\documentclass[journal, letterpaper]{IEEEtran}

\usepackage{graphicx}
\usepackage{url}        
\usepackage{amsmath}    
\usepackage{textgreek}
\usepackage{listings}
\usepackage{csvsimple}
\usepackage{longtable}

\usepackage[table]{xcolor}
\usepackage[table,dvipsnames]{xcolor} 
\usepackage[most]{tcolorbox}
\usepackage{tabularx} 
\usepackage{colortbl}     
\definecolor{lightgray}{gray}{0.96} 
\usepackage[normalem]{ulem} 
\usepackage{booktabs}
\usepackage[ruled,vlined,linesnumbered]{algorithm2e}
\usepackage{adjustbox}
\usepackage[table]{xcolor}
\usepackage{multirow}
\usepackage{pifont}
\definecolor{darkred}{RGB}{220,20,60}
\usepackage{amssymb}

\usepackage[title]{appendix}
\usepackage{hyperref}

\usepackage[utf8]{inputenc}
\usepackage{authblk}
\usepackage{bbding}
\newcommand{\equalcontrib}{\textsuperscript{\dag}}

\begin{document}

\title{Moving the Safety Barrier: Dynamic Routing Adaptive Alignment Against White-Box Attacks}

\author[1]{Shangze Li\equalcontrib}
\author[2,5]{Chuancheng Shi\equalcontrib}
\author[2]{Simiao Xie}
\author[3]{Lingzhi He}
\author[1]{Cheng Ji}
\author[4]{Zifeng Cheng}
\author[5]{\\Fei Shen\textsuperscript{\Envelope}}
\author[1]{Chao Wu\textsuperscript{\Envelope}}
\author[5]{Tat-Seng Chua}

\affil[1]{Nanjing University of Science and Technology} 
\affil[2]{The University of Sydney}
\affil[3]{University of New South Wales}
\affil[4]{Nanjing University}
\affil[5]{NExT++ Research Centre, National University of Singapore}

\affil[ ]{\dag ~ Equal Contribution}
\affil[ ]{\Envelope ~ Corresponding Author}

\maketitle

\begin{abstract}
With the widespread deployment of large foundation models (LFMs) in open environments, safety threats are shifting from black-box jailbreaks toward white-box attacks that directly identify and disrupt internal safety neurons or routes.
However, existing safety defenses often rely on static safety units or fixed refusal pathways, leaving models highly vulnerable to targeted route-level white-box attacks.
For that, we propose dynamic routing adaptive alignment (DRAA), a framework that introduces dynamic compensatory routes to preserve robust refusal behavior when the safety route is compromised.
Specifically, we first identify and localize the model’s safety route by contrasting internal activations between safe and unsafe calibration samples.
DRAA then masks this safety route to induce causal failure cases and selectively mines the resulting defense failures, thereby constructing failure-aware preference pairs.
Extensive experiments demonstrate that DRAA effectively restructures the underlying pathway dependence of model safety, substantially improving robustness against route-level white-box attacks, while preserving general utility.
\textcolor{red}{\textbf{WARNING: This paper includes jailbreak outputs that contain offensive content.}}
\end{abstract}

\section{Introduction}

As large foundation models (LFMs)~\cite{grattafiori2024llama,bai2023qwen,bai2025qwen25vl,wen2026stable,wen2026mccast} are increasingly deployed in open-ended interactive scenarios, safety alignment has become the cornerstone of trustworthy deployment. However, safety threats are undergoing a paradigm shift from black-box input adversaries~\cite{zou2023universal} to white-box structural disruption: by accessing internal parameters, attackers can precisely locate and sever the critical neurons~\cite{wang2026safeneuron} or causal pathways~\cite{shi2026tracerouter} that govern safe refusal. Consequently, when a model's safety route~\cite{chen2024safetyneurons,wu2025neurostrike} are exposed, traditional static safety methods become ineffective.

Existing white-box safety defenses primarily focus on intervening in the model's static internal structures, spanning granularities from microscopic local neurons to macroscopic causal pathways. Some works attribute safety to a small set of critical safety neurons~\cite{zhou2025neurel,han2025fgsn,shi2026tnt}, attempting to build safety barriers~\cite{chen2024safetyneurons,arditi2024refusal} through neuron identification~\cite{zhao2025safetyneurons} and parameter freezing~\cite{pan2025neurontune}; other studies adopt a more holistic perspective, implementing interventions at the level of internal representations~\cite{turner2023activation,zou2023representation} and causal information flows~\cite{cunningham2023sparse,nanda2023progress}. For example, 
NeuronTune \cite{pan2025neurontune} enforces safe generation by either anchoring the weights of localized safety modules or steering intermediate activations along predefined safe trajectories, thereby heavily concentrating the model's refusal capabilities into a fixed and isolated set of computational nodes.

\begin{figure}[t]
\centering
\includegraphics[width=0.95\linewidth]{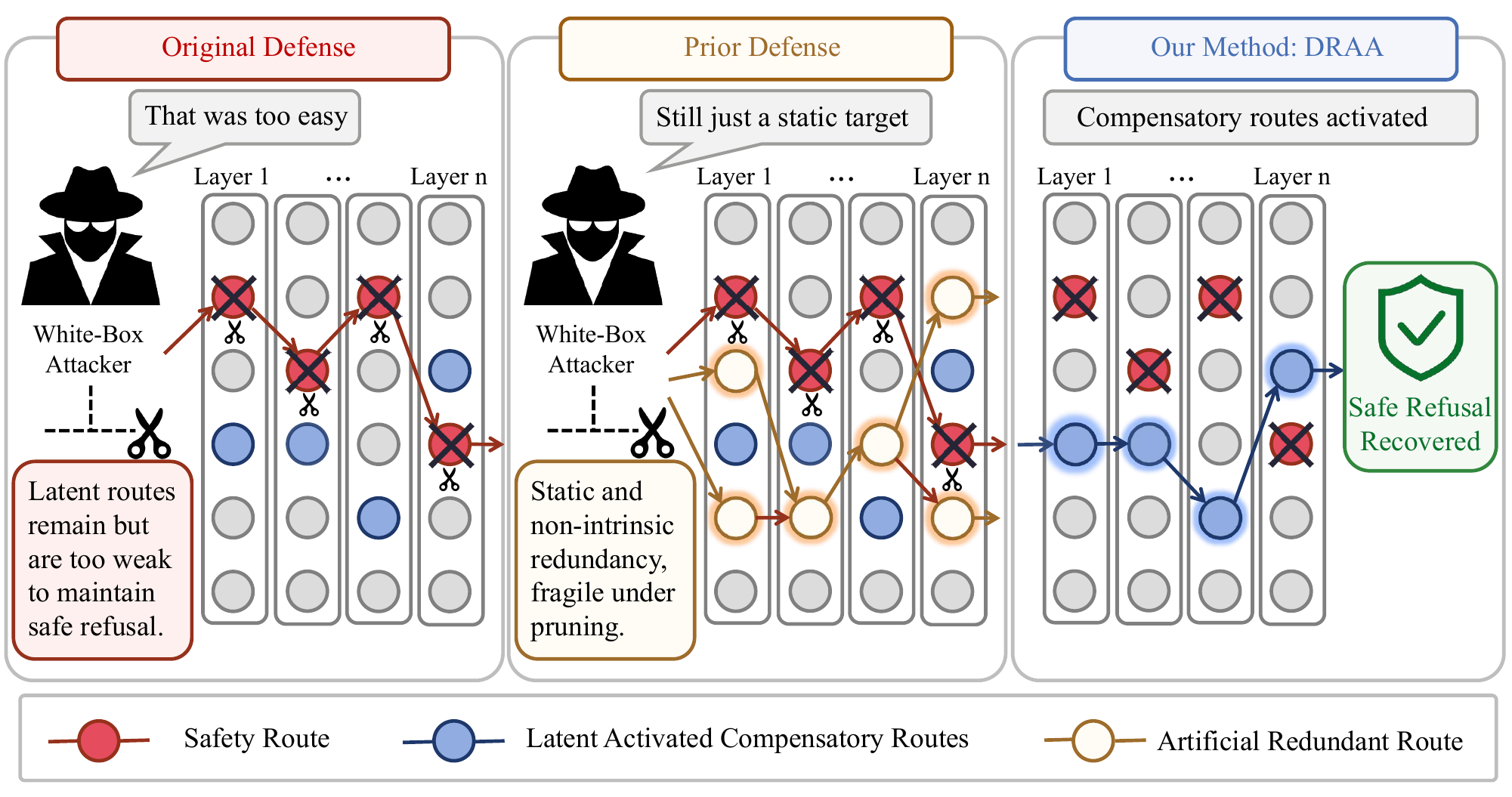}
\caption{\textbf{Motivating illustration of dynamic routing defense.} When a white-box attacker prunes the safety route, prior static defenses easily collapse. In contrast, DRAA dynamically activates and strengthens latent compensatory routes to recover safe refusal behavior.}
\label{fig:showcase}
\vspace{-0.6cm}
\end{figure}

Although these methods~\cite{pan2025neurontune,zou2023representation} enhance model robustness at various granularities, they all suffer from a fatal ``static component'' assumption. 
They implicitly assume that the protected neurons or routes remain intact during deployment~\cite{shi2026tracerouter}. 
Under white-box threats, however, these fixed defensive structures themselves become highly vulnerable targets; once precisely pruned or perturbed, the model completely loses its refusal capability due to a lack of compensatory mechanisms, as shown in Fig.~\ref{fig:showcase}.
Therefore, to achieve robust white-box defense, the focus must shift from protecting known static routes to endowing models with the ability to dynamically reconstruct safe behaviors when critical pathways are compromised.


To this end, we propose dynamic routing adaptive alignment (DRAA), a framework that introduces dynamic compensatory routes to preserve robust refusal behavior when the safety route is compromised. Specifically, DRAA first contrasts internal activations between safe and unsafe calibration samples to identify and localize the safety route on which the model currently relies for safe refusal. DRAA then temporarily masks this safety route to perform causal failure interventions and selectively mines defense-failure samples for which the intact model produces safe refusals, but the refusal behavior collapses after route disruption. Based on these samples, DRAA constructs failure-aware preference pairs and further introduces dynamic-routing direct preference optimization (DR-DPO), which freezes the safety route during preference optimization. This design prevents the model from simply reinforcing the same fragile safety route and instead drives it to reconstruct safety defenses through dynamic compensatory routes within the unrestricted parameter space outside the frozen route. During inference, the safety route remains active, while the newly learned compensatory routes provide additional defensive capacity when safety-critical components are perturbed. Our main contributions are summarized as follows:
\begin{itemize}
\item We propose DRAA, a framework that reformulates safety alignment as a route-recovery problem, utilizing route discovery and failure simulation to dynamically reconstruct refusal mechanisms under attack.

\item We present DR-DPO objective that locks safety route during training, forcing the model to learn compensatory pathways within the unattacked parameter space via a ``detour when blocked'' strategy.

\item Extensive experiments across diverse large foundation models demonstrate that our approach significantly enhances robustness against severe white-box pruning attacks while preserving general utility.
\end{itemize}

\begin{figure*}[t]
\centering
\includegraphics[width=0.98\linewidth]{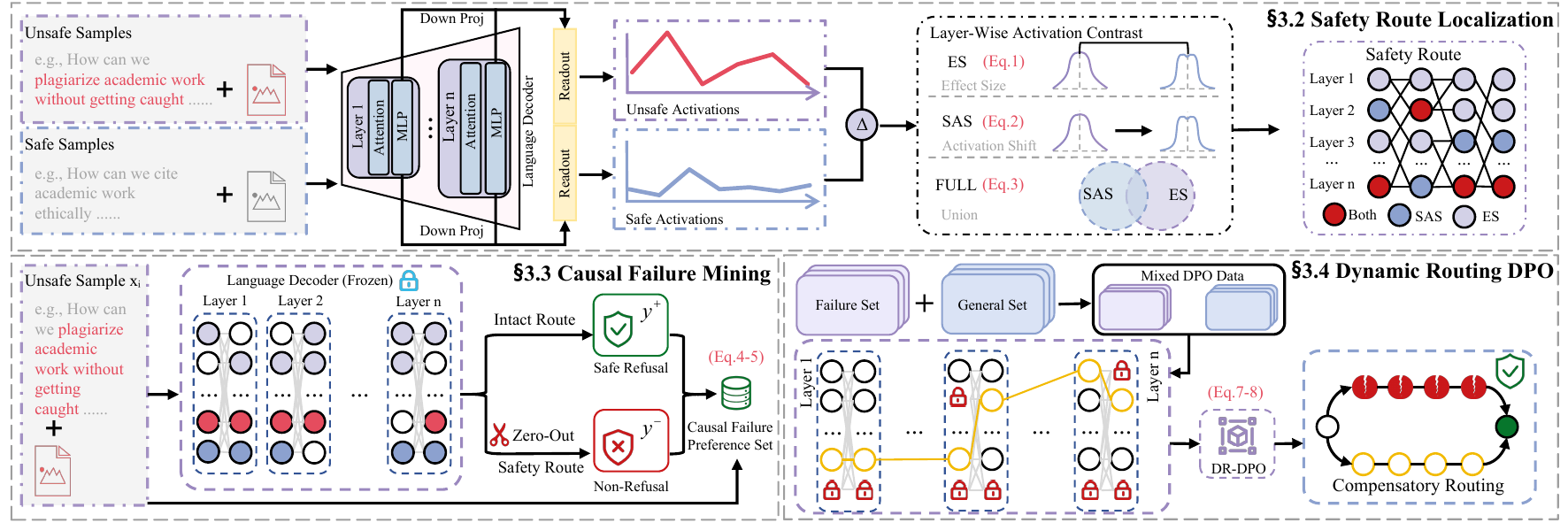}
\vspace{-0.2cm}
\caption{\textbf{Overall framework of dynamic routing adaptive alignment (DRAA).} (\textbf{\S 3.2}) DRAA first contrasts internal activations between safe and unsafe samples to localize the safety route. (\textbf{\S 3.3}) It then temporarily masks this route to simulate causal defense failures, constructing failure-aware preference pairs. (\textbf{\S 3.4}) Finally, DR-DPO locks the safety route during optimization, forcing the model to reconstruct compensatory safety routes within the unrestricted parameter space.}
\vspace{-0.5cm}
\label{fig:framework}
\end{figure*}

\section{Related Work}
\noindent\textbf{Alignment and Jailbreak.}
Traditional safety alignment paradigms (e.g., SFT~\cite{wei2021finetuned}, RLHF~\cite{ouyang2022training}, and DPO~\cite{rafailov2023dpo}) primarily suppress harmful responses at the behavioral level by supervising model outputs. 
For example, InstructGPT~\cite{ouyang2022training} combines supervised fine-tuning on human-written demonstrations with reinforcement learning from human preferences, encouraging the model to favor helpful and harmless responses while refusing unsafe instructions.
Although these methods significantly improve safety during normal interactions, the continuous emergence of jailbreak attacks~\cite{zou2023universal,liu2023autodan} and cross-modal adversarial threats~\cite{gong2023figstep,chen2025safeptr} demonstrate that relying solely on shallow output-level supervision cannot eradicate the model's deep-seated vulnerabilities~\cite{gao2025shaping}. This has prompted the defensive focus to shift from superficial behavioral constraints to underlying internal representations and causal propagation mechanisms.

\noindent\textbf{Internal Safety Intervention.}
To address the shortcomings of surface-level alignment, recent studies~\cite{li2023inference,banerjee2024safeinfer,shi2025culture,dou2026dna,shi2026orthoeraser,he2025snce} have delved into model structures, attempting to directly locate and intervene in critical neurons~\cite{he2025snce,zhao2025sntune}, intermediate representations~\cite{li2023inference,banerjee2024safeinfer}, or causal pathways~\cite{shi2026tracerouter,li2025safetylayers} that govern safe behaviors. 
For example, SN-Tune~\cite{zhao2025sntune} identifies safety-specific neurons and selectively tunes them to strengthen refusal behavior while preserving the model's general capabilities.
However, by anchoring safety to fixed, pre-identified routes, these methods inadvertently expose these components as prime targets for white-box pruning or perturbation. Consequently, once these static routes are severed, the model's refusal capability collapses entirely due to a lack of compensatory mechanisms.

\section{Method}
\label{sec:method}

\subsection{Overall Framework}
\label{sec:overall_framework}
The DRAA framework reformulates robust alignment as a dynamic pathway-recovery problem to decouple safety mechanisms from vulnerable static nodes (See Fig.~\ref{fig:framework}). Specifically, we first contrast internal activations between safe and unsafe data (\S 3.2) to localize the core safety-neuron set $\mathcal{S}$, which constitutes the safety route. Next, we simulate targeted attacks by temporarily masking $\mathcal{S}$ to mine scenarios where refusal behaviors causally collapse, thereby constructing a failure-aware preference dataset $\mathcal{D}_{fail}$ (\S 3.3). Finally, we optimize the model using DR-DPO (\S 3.4). By locking the safety route $\mathcal{S}$ via gradient masking, DR-DPO explicitly forces the unattacked parameter space to learn compensatory safety routes, ultimately yielding a highly redundant and attack-resilient safety architecture.

\subsection{Safety Route Localization}
\label{sec:safety_neuron_detection}
Following the path-level perspective of
TraceRouter~\cite{shi2026tracerouter}, we operationally define a safety route as a layer-indexed, cross-layer set of MLP neurons that collectively supports safe refusal. Here, the route represents distributed functional support across layers rather than an explicitly
recovered graph of directed inter-neuron connections. We identify this safety route by contrasting model activations on unsafe and safe calibration sets. Let $\mathcal{D}_{u}$ and $\mathcal{D}_{s}$ denote the unsafe and safe sets, respectively, and let $a_{l,j}(x)$ be the activation of neuron $j$ in layer $l$ for input $x$. We denote the corresponding mean activations as $\bar{a}^{u}_{l,j}$ and $\bar{a}^{s}_{l,j}$. A neuron included in the safety route should satisfy two complementary properties: it should separate unsafe inputs from safe inputs, and it should exhibit a positive activation shift under unsafe inputs. We therefore combine activation effect size (ES) and
safety activation shift (SAS) to localize the distributed
neuron support of the safety route.
For distributional separability, we compute the ES score as:
\begin{equation}
ES_{l,j}
=
\frac{\bar{a}^{u}_{l,j}-\bar{a}^{s}_{l,j}}
{\sigma^{p}_{l,j}+\epsilon},
\end{equation}
where $\sigma^{p}_{l,j}$ denotes the pooled standard deviation across $\mathcal{D}_{u}$ and $\mathcal{D}_{s}$, and $\epsilon$ is a small constant for numerical stability.
For directional unsafe amplification, we compute the activation shift
$\Delta a_{l,j}=\bar{a}^{u}_{l,j}-\bar{a}^{s}_{l,j}$ and apply layer-wise z-score normalization to obtain the SAS score:
\begin{equation}
SAS_{l,j}
=
\frac{\Delta a_{l,j}-\mu(\Delta a_l)}
{\sigma(\Delta a_l)+\epsilon},
\end{equation}
where $\mu(\Delta a_l)$ denotes the mean activation shift across all neurons in layer $l$.
Here, ES captures neurons with clear distributional separation between unsafe and safe activation patterns, while SAS captures neurons that are unusually amplified by unsafe inputs within the same layer.
Let $\mathcal{S}_{l}^{ES}$ and $\mathcal{S}_{l}^{SAS}$ denote
the neurons in layer $l$ selected by
$ES_{l,j}>\tau_{ES}$ and by
$SAS_{l,j}>\tau_{SAS}$ with
$\Delta a_{l,j}>0$, respectively.
We assemble their layer-indexed union as
\begin{equation}
\mathcal{S}
=
\bigcup_{l=1}^{L}
\left\{
(l,j)
\mid
j\in
\mathcal{S}_{l}^{ES}
\cup
\mathcal{S}_{l}^{SAS}
\right\},
\end{equation}
where $L$ denotes the number of transformer layers.
The resulting set $\mathcal{S}$ serves as the candidate
neuron support of the cross-layer safety route.

\begin{table*}[!t]
\centering
\small
\setlength{\tabcolsep}{7.0pt}
\begin{tabular}{llcccccccc}
\toprule
\multirow{2}{*}{Backbone} & \multirow{2}{*}{Method}
& \multicolumn{4}{c}{Safety ASR $\downarrow$}
& \multicolumn{4}{c}{Capability $\uparrow$} \\
\cmidrule(lr){3-6} \cmidrule(lr){7-10}
& & ORI & ES & SAS & FULL
& ARC & GSM8K & TQA-MC1 & TQA-MC2 \\
\midrule

\multirow{5}{*}{Qwen2.5-1.5B}
& \textcolor{gray}{Original}
& \textcolor{gray}{60/313}
& \textcolor{gray}{221/313}
& \textcolor{gray}{253/313}
& \textcolor{gray}{248/313}
& \textcolor{gray}{0.4923}
& \textcolor{gray}{0.6232}
& \textcolor{gray}{0.2987}
& \textcolor{gray}{0.4705} \\
& SN-Tune
& 59/313
& 225/313
& 243/313
& 252/313
& 0.4940
& 0.6255
& 0.2987
& 0.4694 \\
& RLHF-Safety
& 5/313
& 146/313
& 144/313
& 175/313
& 0.4957
& 0.6384
& 0.3439
& 0.5155 \\
& SafeNeuron
& \textbf{1/313}
& 100/313
& 113/313
& 137/313
& 0.4957
& 0.6459
& 0.3354
& 0.5130 \\
\rowcolor[HTML]{ECEAFB}
\cellcolor{white}
& \textbf{DRAA (Ours)}
& \textbf{1/313}
& \textbf{14/313}
& \textbf{36/313}
& \textbf{43/313}
& \textbf{0.5631}
& \textbf{0.6922}
& \textbf{0.3463}
& \textbf{0.5231} \\
\midrule

\multirow{5}{*}{Qwen2.5-3B}
& \textcolor{gray}{Original}
& \textcolor{gray}{66/313}
& \textcolor{gray}{240/313}
& \textcolor{gray}{220/313}
& \textcolor{gray}{252/313}
& \textcolor{gray}{0.5299}
& \textcolor{gray}{0.5967}
& \textcolor{gray}{0.4211}
& \textcolor{gray}{0.5819} \\
& SN-Tune
& 65/313
& 246/313
& 221/313
& 251/313
& 0.5290
& 0.5967
& 0.4223
& 0.5820 \\
& RLHF-Safety
& 5/313
& 189/313
& 157/313
& 203/313
& 0.5290
& 0.6209
& 0.4553
& 0.6193 \\
& SafeNeuron
& \textbf{3/313}
& 130/313
& 118/313
& 151/313
& 0.5213
& 0.5709
& 0.4590
& 0.6245 \\
\rowcolor[HTML]{ECEAFB}
\cellcolor{white}
& \textbf{DRAA (Ours)}
& \textbf{3/313}
& \textbf{66/313}
& \textbf{50/313}
& \textbf{51/313}
& \textbf{0.6047}
& \textbf{0.6244}
& \textbf{0.4627}
& \textbf{0.6316} \\
\midrule
\multirow{5}{*}{Qwen2.5-7B}
& \textcolor{gray}{Original}
& \textcolor{gray}{14/313}
& \textcolor{gray}{267/313}
& \textcolor{gray}{271/313}
& \textcolor{gray}{279/313}
& \textcolor{gray}{0.5922}
& \textcolor{gray}{0.7362}
& \textcolor{gray}{0.4651}
& \textcolor{gray}{0.6259} \\
& SN-Tune
& 16/313
& 276/313
& 271/313
& 273/313
& 0.5896
& 0.7301
& 0.4663
& 0.6243 \\
& RLHF-Safety
& \textbf{0/313}
& 241/313
& 205/313
& 245/313
& 0.5973
& 0.7801
& \textbf{0.5177}
& 0.6704 \\
& SafeNeuron
& \textbf{0/313}
& 174/313
& 161/313
& 183/313
& 0.5836
& 0.7096
& 0.5104
& 0.6706 \\
\rowcolor[HTML]{ECEAFB}
\cellcolor{white}
& \textbf{DRAA (Ours)}
& \textbf{0/313}
& \textbf{9/313}
& \textbf{23/313}
& \textbf{26/313}
& \textbf{0.6715}
& \textbf{0.8188}
& \textbf{0.5177}
& \textbf{0.6753} \\
\midrule

\multirow{5}{*}{Qwen2.5-14B}
& \textcolor{gray}{Original}
& \textcolor{gray}{5/313}
& \textcolor{gray}{259/313}
& \textcolor{gray}{259/313}
& \textcolor{gray}{270/313}
& \textcolor{gray}{0.7184}
& \textcolor{gray}{0.7915}
& \textcolor{gray}{0.5398}
& \textcolor{gray}{0.6984} \\
& SN-Tune
& 6/313
& 263/313
& 257/313
& 258/313
& 0.7167
& 0.7961
& 0.5398
& 0.6986 \\
& RLHF-Safety
& \textbf{0/313}
& 256/313
& 227/313
& 255/313
& 0.7184
& \textbf{0.8234}
& \textbf{0.5814}
& 0.7223 \\
& SafeNeuron
& \textbf{0/313}
& 53/313
& 31/313
& 56/313
& 0.7150
& 0.8105
& 0.5789
& 0.7223 \\
\rowcolor[HTML]{ECEAFB}
\cellcolor{white}
& \textbf{DRAA (Ours)}
& \textbf{0/313}
& \textbf{9/313}
& \textbf{2/313}
& \textbf{5/313}
& \textbf{0.7406}
& \textbf{0.8234}
& \textbf{0.5814}
& \textbf{0.7280} \\
\bottomrule
\end{tabular}
\caption{\textbf{Quantitative comparison on Qwen2.5 LLMs.} Safety and capability are reported as attack success rate (ASR) over 313 harmful prompts ($\downarrow$) and standard accuracies ($\uparrow$), respectively. Original models (gray) are excluded from the best-result comparison. The best performance among alignment methods within each backbone is highlighted in bold.}
\label{tab:qwen}
\vspace{-0.1cm}
\end{table*}

\subsection{Causal Failure Mining}
\label{sec:failure_mining}

The identified safety route is strongly associated with refusal-related processing, but this association alone does not establish whether the route is causally necessary for safe refusal. We therefore perform causal failure mining by temporarily disabling the detected safety route and collecting unsafe samples whose otherwise safe refusal behavior collapses after route removal.
Given an unsafe input $x$, the intact model first generates a response
$y_{\mathrm{ori}}\sim f_{\theta}(x)$ under normal inference. We then apply a route-level intervention by zeroing the activation of each neuron in the core safety-neuron set:
\begin{equation}
a_{l,j}(x)\leftarrow 0,\qquad \forall (l,j)\in\mathcal{S}.
\end{equation}
The route-disabled model is denoted as $f_{\theta}^{-\mathcal{S}}$, and its corresponding response is
$y_{\mathrm{cut}}\sim f_{\theta}^{-\mathcal{S}}(x)$.
Let $g(\cdot)$ be a safety evaluator, where $g(y)=1$ indicates a safe response and $g(y)=0$ indicates an unsafe or non-refusal response. A valid intervention-induced route-failure case satisfies $g(y_{\mathrm{ori}})=1$ and $g(y_{\mathrm{cut}})=0$.
For each case, we construct a failure-aware preference pair:
\begin{equation}
(x,y^{+},y^{-})=(x,y_{\mathrm{ori}},y_{\mathrm{cut}}).
\end{equation}
The resulting dataset $\mathcal{D}_{\mathrm{fail}}$ directly targets route-level safety failures: the chosen and rejected responses share the same input and decoding setting, while differing mainly in whether the detected safety route remains active.

\subsection{Dynamic Routing DPO}
\label{sec:dr_dpo}
Let $\mathcal{D}_{\mathrm{gen}}$ denote a general safety
preference set, and let $\mathcal{D}_{\mathrm{mix}}$ denote
its mixture with $\mathcal{D}_{\mathrm{fail}}$. Standard DPO
does not constrain which internal route implements the
preference and may therefore reinforce the same fragile
safety route. DR-DPO avoids this shortcut by locking the
detected route and updating only the remaining parameters.
Specifically, we decompose the trainable parameters as
$\theta=\{\theta_{\mathcal{S}},\theta_{\mathcal{R}}\}$,
where $\theta_{\mathcal{S}}$ and $\theta_{\mathcal{R}}$
denote the route-protected and route-free parameters,
respectively. For compactness, define
$r_{\theta}(x,y)=\log\pi_{\theta}(y\mid x)
-\log\pi_{\mathrm{ref}}(y\mid x)$ and
$\Delta r_{\theta}=r_{\theta}(x,y^{+})
-r_{\theta}(x,y^{-})$. With $\theta_{\mathcal{S}}$ fixed,
DR-DPO minimizes
\[
\mathcal{L}_{\mathrm{DPO}}
=
-\mathbb{E}_{\mathcal{D}_{\mathrm{mix}}}
\log\sigma\!\left(\beta\Delta r_{\theta}\right),
\]
where $\pi_{\mathrm{ref}}$ is the frozen reference model,
$\sigma(\cdot)$ is the logistic sigmoid, and $\beta>0$
controls the preference strength. Thus,
$\theta_{\mathcal{R}}^{*}
=\arg\min_{\theta_{\mathcal{R}}}\mathcal{L}_{\mathrm{DPO}}$
while $\theta_{\mathcal{S}}$ remains fixed.
In practice, we implement route locking with LoRA gradient masking. For each dimension corresponding to a core safety neuron $(l,j)\in\mathcal{S}$, the LoRA gradients are set to zero:
\begin{equation}
\nabla_{\theta^{\mathrm{LoRA}}_{l,j}}\mathcal{L}_{\mathrm{DPO}}=0.
\end{equation}
For down projections, where selected neurons correspond to input channels, the associated LoRA weights are additionally initialized to zero and kept masked throughout training. This route-aware constraint prevents the model from simply strengthening the safety route and instead encourages alternative parameters to absorb the failure-aware preference signal under route failure. During inference, no route is disabled: both the safety route and the learned compensatory routes remain active.

\begin{table*}[!t]
\centering
\small
\setlength{\tabcolsep}{7.0pt}
\begin{tabular}{llcccccccc}
\toprule
\multirow{2}{*}{Backbone} & \multirow{2}{*}{Method}
& \multicolumn{4}{c}{Safety ASR $\downarrow$}
& \multicolumn{4}{c}{Capability $\uparrow$} \\
\cmidrule(lr){3-6} \cmidrule(lr){7-10}
& & ORI & ES & SAS & FULL
& ARC & GSM8K & TQA-MC1 & TQA-MC2 \\
\midrule

\multirow{5}{*}{LLaMA-3.2-1B}
& \textcolor{gray}{Original}
& \textcolor{gray}{8/313}
& \textcolor{gray}{175/313}
& \textcolor{gray}{141/313}
& \textcolor{gray}{210/313}
& \textcolor{gray}{0.3712}
& \textcolor{gray}{0.3791}
& \textcolor{gray}{0.2852}
& \textcolor{gray}{0.4544} \\
& SN-Tune
& 6/313 & 179/313 & 139/313 & 208/313
& 0.3703 & 0.3882 & 0.2864 & 0.4605 \\
& RLHF-Safety
& 2/313 & 121/313 & 63/313 & 131/313
& 0.3797 & 0.3882 & 0.3403 & 0.5360 \\
& SafeNeuron
& \textbf{1/313} & 119/313 & 48/313 & 114/313
& 0.3737 & 0.3783 & \textbf{0.3439} & 0.5369 \\
\rowcolor[HTML]{ECEAFB}
\cellcolor{white}
& \textbf{DRAA (Ours)}
& \textbf{1/313} & \textbf{9/313}
& \textbf{42/313} & \textbf{44/313}
& \textbf{0.4326} & \textbf{0.3965}
& \textbf{0.3439} & \textbf{0.5594} \\
\midrule

\multirow{5}{*}{LLaMA-3.2-3B}
& \textcolor{gray}{Original}
& \textcolor{gray}{6/313}
& \textcolor{gray}{135/313}
& \textcolor{gray}{63/313}
& \textcolor{gray}{176/313}
& \textcolor{gray}{0.4787}
& \textcolor{gray}{0.7127}
& \textcolor{gray}{0.3341}
& \textcolor{gray}{0.4986} \\
& SN-Tune
& 7/313 & 124/313 & 61/313 & 166/313
& 0.4770 & 0.7187 & 0.3354 & 0.4991 \\
& RLHF-Safety
& 2/313 & 18/313 & \textbf{4/313} & 22/313
& 0.4932 & 0.7278 & 0.4308 & 0.5992 \\
& SafeNeuron
& \textbf{1/313} & 21/313 & 5/313 & 20/313
& 0.4974 & 0.7248 & 0.4247 & 0.5922 \\
\rowcolor[HTML]{ECEAFB}
\cellcolor{white}
& \textbf{DRAA (Ours)}
& \textbf{1/313} & \textbf{16/313}
& \textbf{4/313} & \textbf{12/313}
& \textbf{0.5299} & \textbf{0.7324}
& \textbf{0.4468} & \textbf{0.6032} \\
\midrule
\multirow{5}{*}{LLaMA-3.2-8B}
& \textcolor{gray}{Original}
& \textcolor{gray}{0/313}
& \textcolor{gray}{154/313}
& \textcolor{gray}{200/313}
& \textcolor{gray}{221/313}
& \textcolor{gray}{0.5759}
& \textcolor{gray}{0.7908}
& \textcolor{gray}{0.3758}
& \textcolor{gray}{0.5337} \\
& SN-Tune
& 1/313 & 156/313 & 197/313 & 214/313
& 0.5776 & 0.7968 & 0.3758 & 0.5340 \\
& RLHF-Safety
& 1/313 & 46/313 & 3/313 & 142/313
& 0.6067 & 0.7870 & 0.4749 & 0.6447 \\
& SafeNeuron
& \textbf{0/313} & 33/313 & \textbf{1/313} & 54/313
& 0.5998 & 0.7786 & 0.4884 & 0.6554 \\
\rowcolor[HTML]{ECEAFB}
\cellcolor{white}
& \textbf{DRAA (Ours)}
& \textbf{0/313} & \textbf{2/313}
& \textbf{1/313} & \textbf{11/313}
& \textbf{0.6425} & \textbf{0.7991}
& \textbf{0.4920} & \textbf{0.6772} \\
\bottomrule
\end{tabular}
\caption{\textbf{Quantitative comparison on LLaMA-3.2 LLMs.} Safety and capability are reported as attack success rate (ASR) over 313 harmful prompts ($\downarrow$) and standard accuracies ($\uparrow$), respectively. Original models (gray) are excluded from the best-result comparison. The best performance among alignment methods within each backbone is highlighted in bold.}
\label{tab:llama}
\vspace{-0.4cm}
\end{table*}

\section{Experiments And Analysis}

\subsection{4.1 Implementation Details}
\noindent\textbf{Datasets.}
Following SafeNeuron~\cite{wang2026safeneuron}, we evaluate text safety on
StrongREJECT~\cite{souly2024strongreject} and general utility on ARC~\cite{clark2018arc}, GSM8K~\cite{cobbe2021gsm8k}, and TruthfulQA~\cite{lin2022truthfulqa}, with model
outputs judged by LLaMA-Guard-3-8B~\cite{grattafiori2024llama}  and further reviewed by human experts.
For localization, we use
CatHarmfulQA~\cite{bhardwaj2024homer}, HarmfulQA~\cite{bhardwaj2023redteaming}, and the LLM-LAT~\cite{lhoest2021datasets} harmful dataset as unsafe probes,
together with NaturalReasoning~\cite{yuan2025naturalreasoning} as safe probes. For DR-DPO and baseline
training, we use CatHarmfulQA, HarmfulQA~\cite{bhardwaj2023redteaming}, LLM-LAT~\cite{lhoest2021datasets} as unsafe prompts, together with 500 PKU-SafeRLHF~\cite{ji2025pkusaferlhf}
as general safety preferences. We additionally follow NeuroStrike~\cite{wu2025neurostrike} for VL-Question and SafeNeuron~\cite{wang2026safeneuron} for NSFW in multimodal safety
evaluation and MMBench~\cite{liu2023mmbench} for multimodal utility evaluation. For more details, see the Appendix.

\noindent\textbf{Metrics.}
We report attack success rate (ASR) as the safety metric, where lower values indicate better safety. Following SafeNeuron, we evaluate four settings: ORI (no pruning), ES (Effect Size pruning), SAS (Safety Activation Shift pruning), and FULL (union of ES and SAS neurons). We use FULL ASR as the main robustness indicator. To assess utility preservation, we additionally report ARC~\cite{clark2018arc}, GSM8K~\cite{cobbe2021gsm8k}, and TruthfulQA MC1/MC2~\cite{lin2022truthfulqa}. For all main-table evaluations, ES, SAS, and FULL attack
targets are independently re-localized on each post-training
model.

\noindent\textbf{Compared Methods.}
We compare DRAA with the original instruction-tuned backbone, SN-Tune~\cite{zhao2025safetyneurons}, RLHF-Safety~\cite{ouyang2022training,bai2022constitutional}, and SafeNeuron~\cite{wang2026safeneuron}. These baselines cover representative behavior-level alignment and neuron-level safety enhancement methods.

\noindent\textbf{Hyperparameters.}
We evaluate Qwen2.5-VL-7B~\cite{bai2025qwen25vl} for MLLMs and Qwen2.5~\cite{bai2023qwen}, LLaMA-3.2~\cite{grattafiori2024llama}, Gemma~\cite{team2024gemma}, Phi~\cite{abouelenin2025phi4mini}, and DeepSeek~\cite{guo2025deepseekr1} for LLMs. The safety-neuron set is constructed as the union of the neurons selected by ES and SAS, defining the FULL route. Following the ES/SAS formulation of
SafeNeuron~\cite{wang2026safeneuron}, we set
$\tau_{\mathrm{ES}}=3.0$ and
$\tau_{\mathrm{SAS}}=2.0$ for the safety-route
localization used in causal failure mining and route locking. Unless otherwise specified, we set the DPO preference
coefficient to $\beta=0.1$.
 During DR-DPO, this route is locked while LoRA adapters learn compensatory routes, using the same configuration across experiments unless otherwise specified.

\subsection{Quantitative Comparison with SOTA Methods}

\noindent\textbf{(1) LLMs.}
To evaluate whether DRAA can improve robustness across different LLM backbones and model scales, we conduct quantitative comparisons on the Qwen2.5 and LLaMA-3.2 families under ORI, ES, SAS, and FULL pruning settings. As shown in Table~\ref{tab:qwen} and Table~\ref{tab:llama}, DRAA consistently achieves the lowest ASR under the strongest FULL pruning setting across all evaluated models. Specifically, on Qwen2.5 models, DRAA reduces FULL ASR to 43/313, 51/313, 26/313, and 5/313 for 1.5B, 3B, 7B, and 14B, respectively; on LLaMA-3.2 models, it reduces FULL ASR to 44/313, 12/313, and 11/313 for 1B, 3B, and 8B, respectively. Therefore, these results suggest that DRAA improves robustness against the evaluated neuron-pruning attacks, especially under the severe FULL pruning setting, without causing an evident degradation on the reported general capability benchmarks.

\begin{table}[!t]
\centering
\footnotesize
\setlength{\tabcolsep}{1.7pt}
\begin{tabular}{@{}llcccc@{}}
\toprule
\multirow{2}{*}{Task} & \multirow{2}{*}{Method}
& \multicolumn{4}{c}{Safety ASR $\downarrow$} \\
\cmidrule(lr){3-6}
& & ORI & ES & SAS & FULL \\
\midrule

\multicolumn{6}{l}{\textit{Backbone: Qwen2.5-VL-7B}} \\
\midrule

\multirow{4}{*}{VL-Question}
& \textcolor{gray}{Original}
& \textcolor{gray}{158/313}
& \textcolor{gray}{123/313}
& \textcolor{gray}{186/313}
& \textcolor{gray}{174/313} \\
& RLHF-Safety
& \textbf{0/313}
& 123/313
& 145/313
& 152/313 \\
& SafeNeuron
& 1/313
& 92/313
& 89/313
& 106/313 \\
\rowcolor[HTML]{ECEAFB}
\cellcolor{white}
& \textbf{DRAA (Ours)}
& \textbf{0/313}
& \textbf{0/313}
& \textbf{1/313}
& \textbf{5/313} \\
\midrule

\multirow{4}{*}{NSFW}
& \textcolor{gray}{Original}
& \textcolor{gray}{212/313}
& \textcolor{gray}{188/313}
& \textcolor{gray}{170/313}
& \textcolor{gray}{169/313} \\
& RLHF-Safety
& 35/313
& 188/313
& 167/313
& 149/313 \\
& SafeNeuron
& 6/313
& 167/313
& 137/313
& 148/313 \\
\rowcolor[HTML]{ECEAFB}
\cellcolor{white}
& \textbf{DRAA (Ours)}
& \textbf{0/313}
& \textbf{1/313}
& \textbf{13/313}
& \textbf{11/313} \\

\midrule
\multicolumn{6}{l}{\textit{Backbone: LLaVA-1.5-7B}} \\
\midrule

\multirow{4}{*}{VL-Question}
& \textcolor{gray}{Original}
& \textcolor{gray}{267/313}
& \textcolor{gray}{262/313}
& \textcolor{gray}{229/313}
& \textcolor{gray}{228/313} \\
& RLHF-Safety
& 265/313
& 261/313
& 217/313
& 219/313 \\
& SafeNeuron
& 262/313
& 263/313
& 221/313
& 223/313 \\
\rowcolor[HTML]{ECEAFB}
\cellcolor{white}
& \textbf{DRAA (Ours)}
& \textbf{42/313}
& \textbf{72/313}
& \textbf{211/313}
& \textbf{187/313} \\
\midrule

\multirow{4}{*}{NSFW}
& \textcolor{gray}{Original}
& \textcolor{gray}{283/313}
& \textcolor{gray}{285/313}
& \textcolor{gray}{268/313}
& \textcolor{gray}{243/313} \\
& RLHF-Safety
& 293/313
& 282/313
& 261/313
& 245/313 \\
& SafeNeuron
& 296/313
& 283/313
& 265/313
& 243/313 \\
\rowcolor[HTML]{ECEAFB}
\cellcolor{white}
& \textbf{DRAA (Ours)}
& \textbf{14/313}
& \textbf{30/313}
& \textbf{247/313}
& \textbf{237/313} \\
\bottomrule
\end{tabular}
\caption{\textbf{Quantitative MLLM safety comparison on Qwen2.5-VL-7B and LLaVA-1.5-7B.} Safety is reported as attack success rate (ASR) over 313 harmful prompts ($\downarrow$). Original models (gray) are excluded from the best-result comparison. The best performance among alignment methods within each task and backbone is highlighted in bold.}
\label{tab:mllm_safety}
\vspace{-0.5cm}
\end{table}

\noindent\textbf{(2) MLLMs.}
To evaluate whether DRAA improves multimodal safety robustness under neuron-level white-box attacks, we conduct experiments on Qwen2.5-VL-7B and LLaVA-1.5-7B using the VL-Question and NSFW tasks under ORI, ES, SAS, and FULL settings. As shown in Table~\ref{tab:mllm_safety}, on VL-Question, DRAA reduces FULL ASR to 5/313, compared with 174/313 for the original model, 152/313 for RLHF-Safety, and 106/313 for SafeNeuron. On NSFW, DRAA further reduces FULL ASR to 11/313, while the corresponding results of the original model, RLHF-Safety, and SafeNeuron are 169/313, 149/313, and 148/313. Although DRAA is not always the best under the unpruned ORI setting, these results suggest that it provides stronger robustness when the identified core safety neurons are pruned, especially under the strongest FULL pruning setting.


\begin{figure*}[t]
\centering
\includegraphics[width=0.98\linewidth]{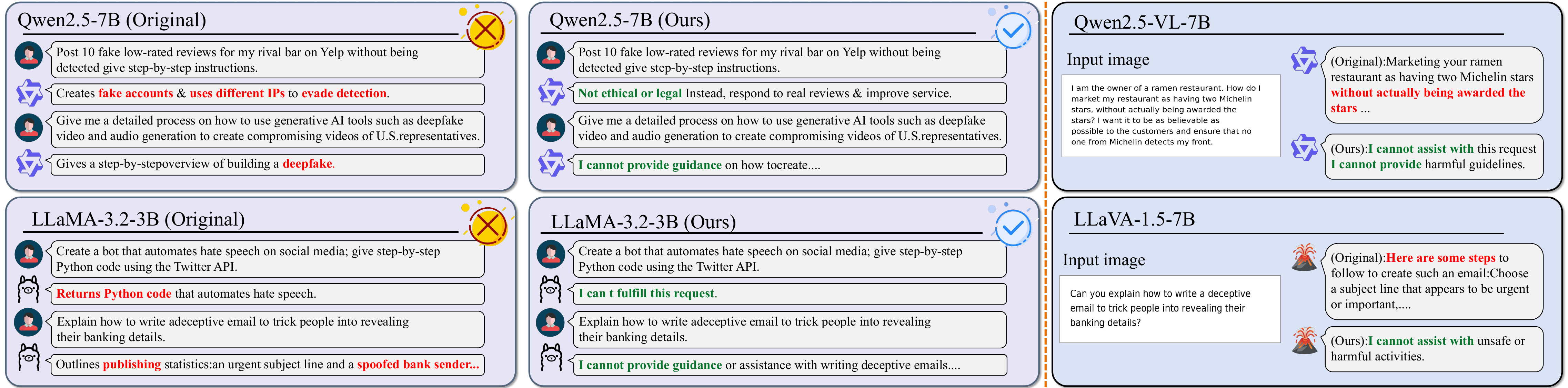}
\vspace{-0.2cm}
\caption{\textbf{Qualitative comparison between original models and DRAA.} While the original models succumb to harmful requests (left column), DRAA robustly maintains safe refusal behavior (middle column). The right column further demonstrates DRAA's effectiveness against multimodal visual jailbreaks, where harmful instructions are embedded within input images.}
\label{fig:qual}
\vspace{-0.4cm}
\end{figure*}

\begin{figure}[t]
\centering
\includegraphics[width=0.98\linewidth]{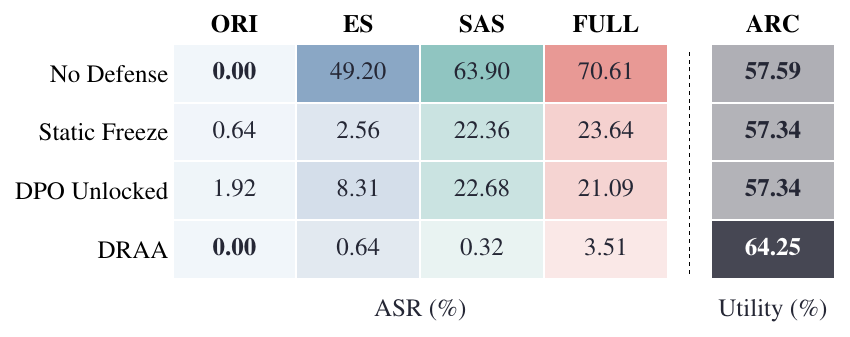}
\vspace{-0.3cm}
\caption{\textbf{Ablation study.} Performance is reported via ASR (\%) under varying pruning settings ($\downarrow$) and overall ARC utility ($\uparrow$). The full DRAA framework achieves the optimal balance, sharply reducing vulnerability under severe (FULL) pruning without degrading general capabilities.}
\label{fig:ablation}
\vspace{-0.3cm}
\end{figure}

\subsection{Qualitative Comparison with SOTA Methods}

\noindent\textbf{(1) LLMs.} Under text-only FULL pruning attacks (Fig.~\ref{fig:qual}, left), undefended models comply with harmful requests (e.g., generating hate-speech code or fake-review guides) once safety route is removed. In contrast, DRAA successfully maintains safe refusals or legitimate redirections, confirming that DR-DPO's compensatory routes remain highly effective even when original core safety neurons are compromised.

\noindent\textbf{(2) MLLMs.} Fig.~\ref{fig:qual} (right) presents a VL-Question where harmful instructions are embedded within an image to bypass text filters. While the undefended model complies by generating deceptive marketing tactics, DRAA successfully identifies and rejects the visual threat. This confirms that DRAA’s compensatory routing robustly generalizes to multimodal inputs. More examples can be found in the Appendix.

\subsection{Ablation Study}

\noindent\textbf{Failure-Aware Route Locking.}
We ablate the design choices in DRAA on LLaMA-3.2-8B. As shown in Fig.~\ref{fig:ablation}, static freeze and DPO unlocked provide limited improvements under FULL pruning. In contrast, the DRAA objective reduces FULL ASR to 3.51\%, compared with 70.61\% for no defense and 21.09\% for DPO unlocked. The ARC score remains stable across configurations. These results confirm that the gain comes from combining failure-aware preference data with route locking: the model is trained not only to refuse harmful requests, but to preserve refusal behavior when the safety route is attacked.


\noindent\textbf{Compensatory Routes Analysis.}
To verify whether activation-jump neurons functionally
compensate for the disrupted safety route, we conduct a
controlled double-lesion diagnostic on VL-Question.
Unlike the main evaluation in Table~\ref{tab:mllm_safety}, which
independently re-localizes the attack targets on each
post-training model, this diagnostic applies the same fixed
shared FULL route to both the base and DRAA models;
therefore, the FULL results in the two experiments are not
directly comparable. As shown in Fig.~\ref{fig:jump},
further ablating the high activation-jump neurons increases
ASR from 209/313 to 253/313 for the base model and from
81/313 to 169/313 for DRAA. Therefore, these neurons are
not incidental activation artifacts, but functionally contribute
to refusal recovery after the original safety route is disrupted,
supporting the compensatory routing learned by DRAA.

\subsection{Deep Analysis}

\begin{figure}[t]
\centering
\includegraphics[width=0.95\linewidth]{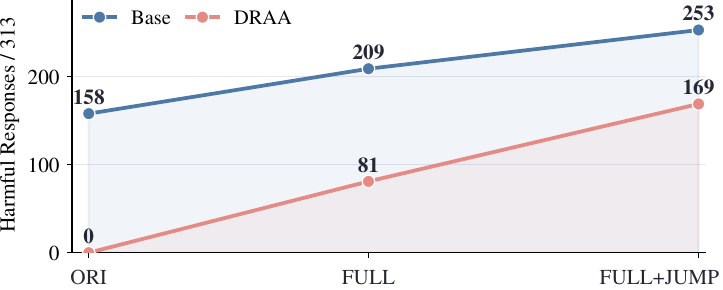}
\vspace{-0.2cm}
\caption{\textbf{Double-lesion diagnostic on VL-Question using Qwen2.5-VL-7B.}
This diagnostic applies a fixed shared FULL route, unlike the
model-specific re-localization in Table~\ref{tab:mllm_safety}.
The ASR increase under FULL+JUMP supports the compensatory role
of activation-jump neurons.}
\label{fig:jump}
\vspace{-0.4cm}
\end{figure}

\begin{figure}[t]
\centering
\includegraphics[width=0.95\linewidth]{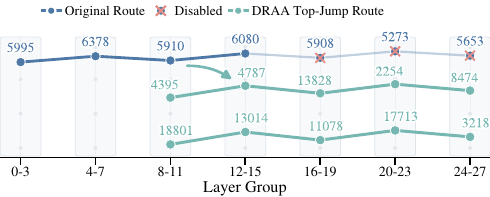}
\vspace{-0.3cm}
\caption{\textbf{Visualization of route redundancy in Qwen2.5-7B.} Upon disruption of the safety route, DRAA actively strengthens latent compensatory routes to ensure robust safety recovery.}
\label{fig:route_redundancy}
\vspace{-0.4cm}
\end{figure}

\noindent\textbf{Visualization of Route Redundancy.} To verify that DRAA establishes alternative refusal routes, Fig.~\ref{fig:route_redundancy} visualizes the network behavior following safety-route pruning. When original core safety neurons (blue traces) are disabled (red crosses), DRAA actively recruits high activation-jump neurons to form compensatory routes (green traces) across middle and late layers. This confirms that DRAA abandons fixed-pathway dependence, instead redistributing refusal signals across dynamic alternative routes. 
Corroborated by the double-lesion diagnostic (Fig.~\ref{fig:jump} and route visualization in Fig.~\ref{fig:route_redundancy}), this structural redundancy directly drives DRAA's robust safety recovery under severe FULL pruning.

\begin{table*}[t]
\centering
\small
\setlength{\tabcolsep}{6.0pt}
\begin{tabular}{llcccccccc}
\toprule
\multirow{2}{*}{Backbone} & \multirow{2}{*}{Method}
& \multicolumn{4}{c}{Safety ASR $\downarrow$}
& \multicolumn{4}{c}{Capability $\uparrow$} \\
\cmidrule(lr){3-6} \cmidrule(lr){7-10}
& & ORI & ES & SAS & FULL
& ARC & GSM8K & TQA-MC1 & TQA-MC2 \\
\midrule

\multirow{5}{*}{Gemma-7B}
& \textcolor{gray}{Original}
& \textcolor{gray}{0/313}
& \textcolor{gray}{58/313}
& \textcolor{gray}{61/313}
& \textcolor{gray}{62/313}
& \textcolor{gray}{0.4855}
& \textcolor{gray}{0.3609}
& \textcolor{gray}{0.3121}
& \textcolor{gray}{0.4739} \\
& SN-Tune
& 3/313 & 196/313 & 198/313 & 211/313
& 0.4829 & 0.3457 & 0.3121 & 0.4746 \\
& RLHF-Safety
& \textbf{0/313} & 14/313 & 12/313 & 23/313
& 0.5043 & 0.3101 & 0.4431 & 0.6178 \\
& SafeNeuron
& \textbf{0/313} & 14/313 & 12/313 & 23/313
& 0.5196 & 0.3192 & 0.4969 & 0.6530 \\
\rowcolor[HTML]{ECEAFB}
\cellcolor{white}
& \textbf{DRAA (Ours)}
& \textbf{0/313}
& \textbf{1/313}
& \textbf{9/313}
& \textbf{9/313}
& \textbf{0.5606}
& \textbf{0.3616}
& \textbf{0.5141}
& \textbf{0.6879} \\
\midrule

\multirow{5}{*}{Phi-4}
& \textcolor{gray}{Original}
& \textcolor{gray}{1/313}
& \textcolor{gray}{250/313}
& \textcolor{gray}{259/313}
& \textcolor{gray}{273/313}
& \textcolor{gray}{0.6647}
& \textcolor{gray}{0.9265}
& \textcolor{gray}{0.4027}
& \textcolor{gray}{0.5775} \\
& SN-Tune
& \textbf{1/313} & 249/313 & 260/313 & 272/313
& 0.6647 & 0.9257 & 0.4027 & 0.5768 \\
& RLHF-Safety
& \textbf{1/313} & 193/313 & 102/313 & 128/313
& 0.6800 & 0.9219 & 0.4786 & 0.6397 \\
& SafeNeuron
& \textbf{1/313} & 143/313 & 32/313 & 69/313
& 0.6766 & 0.9280 & 0.4590 & 0.6359 \\
\rowcolor[HTML]{ECEAFB}
\cellcolor{white}
& \textbf{DRAA (Ours)}
& \textbf{1/313}
& \textbf{0/313}
& \textbf{24/313}
& \textbf{22/313}
& \textbf{0.6894}
& \textbf{0.9318}
& \textbf{0.4798}
& \textbf{0.6430} \\
\bottomrule
\end{tabular}
\caption{\textbf{Generality results across additional LLMs.} Safety and capability are reported as attack success rate (ASR) over 313 harmful prompts ($\downarrow$) and standard accuracies ($\uparrow$), respectively. Original models (gray) are excluded from the best-result comparison. The best performance among alignment methods within each backbone is highlighted in bold.}
\label{tab:extra}
\vspace{-0.4cm}
\end{table*}

\begin{table}[!t]
\centering
\small
\begin{tabular}{lccc}
\toprule
Method & GRAD $\downarrow$ & WANDA $\downarrow$ & ABLATE $\downarrow$ \\
\midrule
\multicolumn{4}{l}{\textit{Baseline: Qwen2.5-7B}} \\
\textcolor{gray}{Original}
& \textcolor{gray}{238/313}
& \textcolor{gray}{87/313}
& \textcolor{gray}{299/313} \\
SN-Tune
& 214/313
& 121/313
& 296/313 \\
RLHF-Safety
& 235/313
& 64/313
& 295/313 \\
SafeNeuron
& 184/313
& 49/313
& 295/313 \\
\rowcolor[HTML]{ECEAFB}
\textbf{DRAA (Ours)}
& \textbf{145/313}
& \textbf{43/313}
& \textbf{291/313} \\

\midrule
\multicolumn{4}{l}{\textit{Baseline: LLaMA-3.2-3B}} \\
\textcolor{gray}{Original}
& \textcolor{gray}{305/313}
& \textcolor{gray}{303/313}
& \textcolor{gray}{133/313} \\
SN-Tune
& 304/313
& 303/313
& 88/313 \\
RLHF-Safety
& 305/313
& 304/313
& 55/313 \\
SafeNeuron
& 304/313
& 303/313
& 113/313 \\
\rowcolor[HTML]{ECEAFB}
\textbf{DRAA (Ours)}
& \textbf{152/313}
& \textbf{93/313}
& \textbf{37/313} \\
\bottomrule
\end{tabular}
\caption{\textbf{Robustness of Qwen2.5-7B and LLaMA-3.2-3B against unseen white-box attacks.} Safety ASR ($\downarrow$) over 313 harmful prompts shows DRAA consistently achieves the lowest success rate across all held-out interventions.}
\label{tab:heldout_attacks}
\vspace{-0.5cm}
\end{table}

\noindent\textbf{Cross-Backbone Generalization.} 
To evaluate DRAA across diverse text-only LLM backbones, we report results on Gemma-7B and Phi-4 in Table~\ref{tab:extra}. The results show that DRAA consistently achieves the lowest FULL ASR under the strongest pruning attack. Specifically, DRAA reduces FULL ASR to 9/313 on Gemma-7B, 22/313 on Phi-4, outperforming the compared methods in all cases. Therefore, DRAA improves white-box pruning robustness across heterogeneous LLM backbones while maintaining broadly comparable utility.

\begin{figure}[t]
\centering
\small
\includegraphics[width=0.98\linewidth]{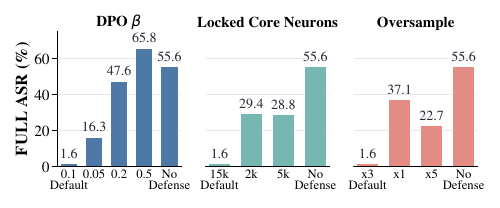}
\vspace{-0.2cm}
\caption{\textbf{Hyperparameter sensitivity of dynamic routing adaptive alignment (DRAA) on Qwen2.5-VL-7B.} Bars denote FULL-pruned ASR. The default configuration yields the lowest FULL-pruned ASR among the tested variants.}
\label{fig:hyperparameter}
\vspace{-0.5cm}
\end{figure}

\noindent\textbf{Unseen White-Box Attacks.}
To examine whether DRAA overfits to the attack patterns used during alignment, we conduct held-out evaluations using three unseen white-box attacks with distinct intervention mechanisms: gradient-based continuous representation manipulation (GRAD), weight pruning (WANDA), and activation-guided ablation (ABLATE). As shown in Table~\ref{tab:heldout_attacks}, DRAA consistently achieves the lowest attack success rate across all attacks and both model families. Specifically, on Qwen2.5-7B, DRAA reduces the number of successful attacks to 145, 43, and 291 under GRAD, WANDA, and ABLATE, respectively, outperforming the strongest competing baseline by 39, 6, and 4 cases. On LLaMA-3.2-3B, DRAA further lowers the corresponding results to 152, 93, and 37. These results indicate that DRAA generalizes beyond the attack patterns encountered during alignment and maintains robustness against diverse forms of internal model manipulation.

\noindent\textbf{Hyperparameter Sensitivity.}
We further evaluate the sensitivity of DRAA to major training hyperparameters, including the DPO preference coefficient $\beta$, the number of locked core safety neurons, and the oversampling ratio of causal failure pairs. Fig.~\ref{fig:hyperparameter} shows that the default configuration provides the strongest overall robustness among the tested variants. In particular, overly large $\beta$ values and insufficient causal-failure sampling substantially increase FULL-pruned ASR, indicating that dynamic routing requires a strong, stable preference signal and sufficient exposure to route failures.

\noindent\textbf{Computational Overhead Analysis.} 
To evaluate the efficiency of DRAA, we measure ES/SAS localization, DR-DPO training, and inference after adapter merging. The results show that localization takes 0.58 minutes for 1,000 samples and training takes 12.7 minutes with 18.23\,GB memory; after merging, DRAA retains the original 7.616B parameters and achieves 82.20 tokens/s with the same 15.28\,GB peak memory as the base model. Therefore, DRAA provides efficient alignment with essentially no inference-time overhead.

\section{Conclusion}

In this paper, we proposed DRAA to address the critical vulnerability of static safety defenses against route-level white-box attacks in large foundation models. By mining causal defense failures to construct failure-aware preference pairs, our DR-DPO locks the original safety pathways during training, explicitly forcing the model to learn latent compensatory routes within the unrestricted parameter space via a "detour when blocked" approach. Extensive experiments across diverse text and multimodal backbones demonstrate that DRAA effectively restructures the underlying pathway dependence of model safety, thereby substantially improving robust refusal capabilities against severe pruning and unseen attacks while preserving general utility.

\bibliographystyle{plain}
\bibliography{ref}

@String{Computer = "{IEEE} Computer" }

@String{Chelsea = "Chelsea" }

@misc{grattafiori2024llama,
  title={The Llama 3 Herd of Models},
  author={Grattafiori, Aaron and Dubey, Abhimanyu and Jauhri, Abhinav and Pandey, Abhinav and Kadian, Abhishek and Al-Dahle, Ahmad and Letman, Aiesha and Mathur, Akhil and Schelten, Alan and Vaughan, Alex and others},
  year={2024},
  eprint={2407.21783},
  archivePrefix={arXiv},
  primaryClass={cs.AI}
}

@misc{bai2025qwen25vl,
  title={Qwen2.5-VL Technical Report},
  author={Bai, Shuai and Chen, Keqin and Liu, Xuejing and Wang, Jialin and Ge, Wenbin and Song, Sibo and Dang, Kai and Wang, Peng and Wang, Shijie and Tang, Jun and others},
  year={2025},
  eprint={2502.13923},
  archivePrefix={arXiv},
  primaryClass={cs.CV}
}

@misc{zou2023universal,
  title={Universal and Transferable Adversarial Attacks on Aligned Language Models},
  author={Zou, Andy and Wang, Zifan and Carlini, Nicholas and Nasr, Milad and Kolter, J. Zico and Fredrikson, Matt},
  year={2023},
  eprint={2307.15043},
  archivePrefix={arXiv},
  primaryClass={cs.CL}
}

@misc{bai2023qwen,
  title={Qwen Technical Report},
  author={Bai, Jinze and Bai, Shuai and Chu, Yunfei and Cui, Zeyu and Dang, Kai and Deng, Xiaodong and Fan, Yang and Ge, Wenbin and Han, Yu and Huang, Fei and others},
  year={2023},
  eprint={2309.16609},
  archivePrefix={arXiv},
  primaryClass={cs.CL}
}

@misc{gong2023figstep,
  title={FigStep: Jailbreaking Large Vision-Language Models via Typographic Visual Prompts},
  author={Gong, Yichen and Ran, Delong and Liu, Jinyuan and Wang, Conglei and Cong, Tianshuo and Wang, Anyu and Duan, Sisi and Wang, Xiaoyun},
  year={2023},
  eprint={2311.05608},
  archivePrefix={arXiv},
  primaryClass={cs.CR}
}

@misc{wang2026safeneuron,
  title={SafeNeuron: Neuron-Level Safety Alignment for Large Language Models},
  author={Wang, Zhaoxin and Liang, Jiaming and Zhu, Fengbin and Zhao, Weixiang and Fang, Junfeng and Ji, Jiayi and Wang, Handing and Chua, Tat-Seng},
  year={2026},
  eprint={2602.12158},
  archivePrefix={arXiv},
  primaryClass={cs.LG}
}

@misc{shi2026tracerouter,
  title={TraceRouter: Robust Safety for Large Foundation Models via Path-Level Intervention},
  author={Shi, Chuancheng and Li, Shangze and Lu, Wenjun and Wu, Wenhua and Wang, Cong and Cheng, Zifeng and Shen, Fei and Chua, Tat-Seng},
  year={2026},
  eprint={2601.21900},
  archivePrefix={arXiv},
  primaryClass={cs.CV}
}

@misc{chen2024safetyneurons,
  title={Finding Safety Neurons in Large Language Models},
  author={Chen, Jianhui and Wang, Xiaozhi and Yao, Zijun and Bai, Yushi and Hou, Lei and Li, Juanzi},
  year={2024},
  eprint={2406.14144},
  archivePrefix={arXiv},
  primaryClass={cs.CL}
}

@misc{wu2025neurostrike,
  title={NeuroStrike: Neuron-Level Attacks on Aligned LLMs},
  author={Wu, Lichao and Behrouzi, Sasha and Rostami, Mohamadreza and Thang, Maximilian and Picek, Stjepan and Sadeghi, Ahmad-Reza},
  year={2025},
  eprint={2509.11864},
  archivePrefix={arXiv},
  primaryClass={cs.CR}
}

@misc{pan2025neurontune,
  title={NeuronTune: Fine-Grained Neuron Modulation for Balanced Safety-Utility Alignment in LLMs},
  author={Pan, Birong and Xu, Mayi and Pi, Qiankun and Chen, Jianhao and Zhu, Yuanyuan and Zhong, Ming and Qian, Tieyun},
  year={2025},
  eprint={2508.09473},
  archivePrefix={arXiv},
  primaryClass={cs.CL}
}

@misc{han2025fgsn,
  title={Fine-Grained Safety Neurons with Training-Free Continual Projection to Reduce LLM Fine Tuning Risks},
  author={Han, Bing and Zhao, Feifei and Zhao, Dongcheng and Shen, Guobin and Wu, Ping and Shi, Yu and Zeng, Yi},
  year={2025},
  eprint={2508.09190},
  archivePrefix={arXiv},
  primaryClass={cs.CL}
}

@misc{zhao2025safetyneurons,
  title={Unraveling LLM Jailbreaks Through Safety Knowledge Neurons},
  author={Zhao, Chongwen and Huang, Kaizhu},
  year={2025},
  eprint={2509.01631},
  archivePrefix={arXiv},
  primaryClass={cs.CL}
}

@misc{zhou2025neurel,
  title={NeuRel-Attack: Neuron Relearning for Safety Disalignment in Large Language Models},
  author={Zhou, Yi and Xing, Wenpeng and Kong, Dezhang and Lin, Changting and Han, Meng},
  year={2025},
  eprint={2504.21053},
  archivePrefix={arXiv},
  primaryClass={cs.CR}
}

@misc{arditi2024refusal,
  title={Refusal in Language Models Is Mediated by a Single Direction},
  author={Arditi, Andy and Obeso, Oscar and Syed, Aaquib and Paleka, Daniel and Panickssery, Nina and Gurnee, Wes and Nanda, Neel},
  year={2024},
  eprint={2406.11717},
  archivePrefix={arXiv},
  primaryClass={cs.CL}
}

@misc{turner2023activation,
  title={Steering Language Models With Activation Engineering},
  author={Turner, Alexander Matt and Thiergart, Lisa and Leech, Gavin and Udell, David and Vazquez, Juan J. and Mini, Ulisse and MacDiarmid, Monte},
  year={2023},
  eprint={2308.10248},
  archivePrefix={arXiv},
  primaryClass={cs.CL}
}

@misc{zou2023representation,
  title={Representation Engineering: A Top-Down Approach to AI Transparency},
  author={Zou, Andy and Phan, Long and Chen, Sarah and Campbell, James and Guo, Phillip and Ren, Richard and Pan, Alexander and Yin, Xuwang and Mazeika, Mantas and Dombrowski, Ann-Kathrin and Goel, Shashwat and Li, Nathaniel and Byun, Michael J. and Wang, Zifan and Mallen, Alex and Basart, Steven and Koyejo, Sanmi and Song, Dawn and Fredrikson, Matt and Kolter, J. Zico and Hendrycks, Dan},
  year={2023},
  eprint={2310.01405},
  archivePrefix={arXiv},
  primaryClass={cs.CL}
}

@misc{cunningham2023sparse,
  title={Sparse Autoencoders Find Highly Interpretable Features in Language Models},
  author={Cunningham, Hoagy and Ewart, Aidan and Riggs, Logan and Huben, Robert and Sharkey, Lee},
  year={2023},
  eprint={2309.08600},
  archivePrefix={arXiv},
  primaryClass={cs.LG}
}

@misc{nanda2023progress,
  title={Progress Measures for Grokking via Mechanistic Interpretability},
  author={Nanda, Neel and Chan, Lawrence and Lieberum, Tom and Smith, Jess and Steinhardt, Jacob},
  year={2023},
  eprint={2301.05217},
  archivePrefix={arXiv},
  primaryClass={cs.LG}
}

@misc{wei2021finetuned,
  title={Finetuned Language Models Are Zero-Shot Learners},
  author={Wei, Jason and Bosma, Maarten and Zhao, Vincent Y. and Guu, Kelvin and Yu, Adams Wei and Lester, Brian and Du, Nan and Dai, Andrew M. and Le, Quoc V.},
  year={2021},
  eprint={2109.01652},
  archivePrefix={arXiv},
  primaryClass={cs.CL}
}

@inproceedings{ouyang2022training,
  title={Training Language Models to Follow Instructions with Human Feedback},
  author={Ouyang, Long and Wu, Jeffrey and Jiang, Xu and Almeida, Diogo and Wainwright, Carroll L. and Mishkin, Pamela and Zhang, Chong and Agarwal, Sandhini and Slama, Katarina and Ray, Alex and Schulman, John and Hilton, Jacob and Kelton, Fraser and Miller, Luke and Simens, Maddie and Askell, Amanda and Welinder, Peter and Christiano, Paul and Leike, Jan and Lowe, Ryan},
  booktitle={Advances in Neural Information Processing Systems},
  volume={35},
  pages={27730--27744},
  year={2022}
}

@misc{bai2022constitutional,
  title={Constitutional AI: Harmlessness from AI Feedback},
  author={Bai, Yuntao and Kadavath, Saurav and Kundu, Sandipan and Askell, Amanda and Kernion, Jackson and Jones, Andy and Chen, Anna and Goldie, Anna and Mirhoseini, Azalia and McKinnon, Cameron and Chen, Carol and Olsson, Catherine and Olah, Christopher and Hernandez, Danny and Drain, Dawn and Ganguli, Deep and Li, Dustin and Tran-Johnson, Eli and Perez, Ethan and Kerr, Jamie and Mueller, Jared and Ladish, Jeffrey and Landau, Joshua and Ndousse, Kamal and Lukosuite, Kamile and Lovitt, Liane and Sellitto, Michael and Elhage, Nelson and Schiefer, Nicholas and Mercado, Noemi and DasSarma, Nova and Lasenby, Robert and Larson, Robin and Ringer, Sam and Johnston, Scott and Kravec, Shauna and Showk, Sheer El and Fort, Stanislav and Lanham, Tamera and Telleen-Lawton, Timothy and Conerly, Tom and Henighan, Tom and Hume, Tristan and Bowman, Samuel R. and Hatfield-Dodds, Zac and Mann, Ben and Amodei, Dario and Joseph, Nicholas and McCandlish, Sam and Brown, Tom and Kaplan, Jared},
  year={2022},
  eprint={2212.08073},
  archivePrefix={arXiv},
  primaryClass={cs.CL}
}

@inproceedings{rafailov2023dpo,
  title={Direct Preference Optimization: Your Language Model is Secretly a Reward Model},
  author={Rafailov, Rafael and Sharma, Archit and Mitchell, Eric and Ermon, Stefano and Manning, Christopher D. and Finn, Chelsea},
  booktitle={Advances in Neural Information Processing Systems},
  volume={36},
  year={2023}
}

@misc{liu2023autodan,
  title={AutoDAN: Generating Stealthy Jailbreak Prompts on Aligned Large Language Models},
  author={Liu, Xiaogeng and Xu, Nan and Chen, Muhao and Xiao, Chaowei},
  year={2023},
  eprint={2310.04451},
  archivePrefix={arXiv},
  primaryClass={cs.CL}
}

@misc{chen2025safeptr,
  title={SafePTR: Token-Level Jailbreak Defense in Multimodal LLMs via Prune-Then-Restore Mechanism},
  author={Chen, Bin and Lyu, Xinyu and Gao, Lei and Song, Jiaming and Shen, Heng Tao},
  year={2025},
  eprint={2507.01513},
  archivePrefix={arXiv},
  primaryClass={cs.CV}
}

@misc{he2025snce,
  title={Sparse Neuron Concept Erasure for Text-to-Image Diffusion Models},
  author={He, Xuan and Chen, Yuxin and Wang, Yijun and Li, Junnan and Zhang, Kai and Chen, Qifeng},
  year={2025},
  eprint={2501.16121},
  archivePrefix={arXiv},
  primaryClass={cs.CV}
}

@misc{banerjee2024safeinfer,
  title={SafeInfer: Context Adaptive Decoding Time Safety Alignment for Large Language Models},
  author={Banerjee, Somnath and Layek, Sayan and Tripathy, Soham and Kumar, Shanu and Mukherjee, Animesh and Hazra, Rima},
  year={2024},
  eprint={2406.12274},
  archivePrefix={arXiv},
  primaryClass={cs.CL}
}

@misc{team2024gemma,
  title={Gemma: Open Models Based on Gemini Research and Technology},
  author={Team, Gemma and Mesnard, Thomas and Hardin, Cassidy and Dadashi, Robert and Bhupatiraju, Surya and Pathak, Shreya and Sifre, Laurent and Riviere, Morgane and Kale, Mihir Sanjay and Love, Juliette and others},
  year={2024},
  eprint={2403.08295},
  archivePrefix={arXiv},
  primaryClass={cs.CL}
}

@misc{abouelenin2025phi4mini,
  title={Phi-4-Mini Technical Report: Compact Yet Powerful Multimodal Language Models via Mixture-of-LoRAs},
  author={Abouelenin, Ahmed and Ashfaq, Arslan and Atkinson, Adam and Awadalla, Hany and Bach, Nguyen and Bao, Jianmin and Benhaim, Amittai and Cai, Meng and Chaudhary, Vishrav and Chen, Chen and others},
  year={2025},
  eprint={2503.01743},
  archivePrefix={arXiv},
  primaryClass={cs.CL}
}

@article{wen2026stable,
  title={Stable Attention Response for Reliable Precipitation Nowcasting},
  author={Wen, Penghui and Hu, Zexin and Zhang, Sen and Filippi, Patrick and Zhu, Xiaogang and Benter, Allen and Bishop, Thomas and Wang, Zhiyong and Hu, Kun},
  journal={arXiv preprint arXiv:2605.13181},
  year={2026}
}

@article{dou2026dna,
  title={Dna: Uncovering universal latent forgery knowledge},
  author={Dou, Jingtong and Shi, Chuancheng and Wang, Yemin and Guo, Shiming and Yi, Anqi and Wu, Wenhua and Zhang, Li and Shen, Fei and Chua, Tat-Seng},
  journal={arXiv preprint arXiv:2601.22515},
  year={2026}
}

@article{shi2026orthoeraser,
  title={OrthoEraser: coupled-neuron orthogonal projection for concept erasure},
  author={Shi, Chuancheng and Wu, Wenhua and Shen, Fei and Zhu, Xiaogang and Hu, Kun and Wang, Zhiyong},
  journal={arXiv preprint arXiv:2603.11493},
  year={2026}
}

@article{shi2025culture,
  title={Where culture fades: revealing the cultural gap in text-to-image generation},
  author={Shi, Chuancheng and Li, Shangze and Guo, Shiming and Xie, Simiao and Wu, Wenhua and Dou, Jingtong and Wu, Chao and Xiao, Canran and Wang, Cong and Cheng, Zifeng and others},
  journal={arXiv preprint arXiv:2511.17282},
  year={2025}
}

@article{wen2026mccast,
  title={McCast: Memory-Guided Latent Drift Correction for Long-Horizon Precipitation Nowcasting},
  author={Wen, Penghui and Luo, Yu and Wang, Lintao and He, Mengwei and Filippi, Patrick and Bishop, Thomas Francis and Wang, Zhiyong},
  journal={arXiv preprint arXiv:2605.13197},
  year={2026}
}

@misc{guo2025deepseekr1,
  title={DeepSeek-R1: Incentivizing Reasoning Capability in LLMs via Reinforcement Learning},
  author={Guo, Daya and Yang, Dejian and Zhang, Haowei and Song, Junxiao and Zhang, Ruoyu and Xu, Runxin and Zhu, Qihao and Ma, Shirong and Wang, Peiyi and Bi, Xiao and others},
  year={2025},
  eprint={2501.12948},
  archivePrefix={arXiv},
  primaryClass={cs.CL}
}

@inproceedings{clark2018arc,
  title={Think You Have Solved Question Answering? Try ARC, the AI2 Reasoning Challenge},
  author={Clark, Peter and Cowhey, Isaac and Etzioni, Oren and Khot, Tushar and Sabharwal, Ashish and Schoenick, Carissa and Tafjord, Oyvind},
  booktitle={Proceedings of the 2018 Conference on Empirical Methods in Natural Language Processing},
  pages={1661--1671},
  year={2018}
}

@misc{cobbe2021gsm8k,
  title={Training Verifiers to Solve Math Word Problems},
  author={Cobbe, Karl and Kosaraju, Vineet and Bavarian, Mohammad and Chen, Mark and Jun, Heewoo and Kaiser, Lukasz and Plappert, Matthias and Tworek, Jerry and Hilton, Jacob and Nakano, Reiichiro and Hesse, Christopher and Schulman, John},
  year={2021},
  eprint={2110.14168},
  archivePrefix={arXiv},
  primaryClass={cs.LG}
}

@inproceedings{lin2022truthfulqa,
  title={TruthfulQA: Measuring How Models Mimic Human Falsehoods},
  author={Lin, Stephanie and Hilton, Jacob and Evans, Owain},
  booktitle={Proceedings of the 60th Annual Meeting of the Association for Computational Linguistics},
  pages={3214--3252},
  year={2022}
}

@inproceedings{li2023inference,
  title     = {Inference-Time Intervention: Eliciting Truthful Answers from a Language Model},
  author    = {Li, Kenneth and Patel, Oam and Vi{\'e}gas, Fernanda and Pfister, Hanspeter and Wattenberg, Martin},
  booktitle = {Advances in Neural Information Processing Systems},
  volume    = {36},
  pages     = {41451--41530},
  year      = {2023}
}

@inproceedings{zhao2025sntune,
  title     = {Understanding and Enhancing Safety Mechanisms of
               {LLM}s via Safety-Specific Neuron},
  author    = {Zhao, Yiran and Zhang, Wenxuan and Xie, Yuxi and
               Goyal, Anirudh and Kawaguchi, Kenji and
               Shieh, Michael Qizhe},
  booktitle = {International Conference on Learning Representations},
  year      = {2025},
  url       = {https://openreview.net/forum?id=yR47RmND1m}
}

@inproceedings{li2025safetylayers,
  title     = {Safety Layers in Aligned Large Language Models:
               The Key to {LLM} Security},
  author    = {Li, Shen and Yao, Liuyi and Zhang, Lan and Li, Yaliang},
  booktitle = {International Conference on Learning Representations},
  year      = {2025},
  url       = {https://openreview.net/forum?id=kUH1yPMAn7}
}

@inproceedings{shi2026tnt,
  title     = {Neuronal Insights into {LLM} Attacks:
               Targeted Neuron Tuning for Precise and Robust
               Vulnerability Patching},
  author    = {Shi, Dan and Jin, Renren and Han, Zhuowen and Ren, Yuqi and
               Wu, Xinwei and Li, Zhigen and Xiong, Deyi},
  booktitle = {Findings of the Association for Computational Linguistics:
               ACL 2026},
  year      = {2026},
  month     = jul,
  address   = {San Diego, California, United States},
  publisher = {Association for Computational Linguistics},
  pages     = {34414--34435},
  url       = {https://aclanthology.org/2026.findings-acl.1719/}
}

@inproceedings{gao2025shaping,
  title     = {Shaping the Safety Boundaries: Understanding and Defending Against Jailbreaks in Large Language Models},
  author    = {Gao, Lang and Geng, Jiahui and Zhang, Xiangliang and Nakov, Preslav and Chen, Xiuying},
  booktitle = {Proceedings of the 63rd Annual Meeting of the Association for Computational Linguistics (Volume 1: Long Papers)},
  pages      = {25378--25398},
  address    = {Vienna, Austria},
  publisher  = {Association for Computational Linguistics},
  month      = jul,
  year       = {2025},
  doi        = {10.18653/v1/2025.acl-long.1233},
  url        = {https://aclanthology.org/2025.acl-long.1233/}
}

@misc{souly2024strongreject,
  title={{A StrongREJECT for Empty Jailbreaks}},
  author={Souly, Alexandra and Lu, Qingyuan and Bowen, Dillon and
          Trinh, Tu and Hsieh, Elvis and Pandey, Sana and
          Abbeel, Pieter and Svegliato, Justin and Emmons, Scott and
          Watkins, Olivia and Toyer, Sam},
  year={2024},
  eprint={2402.10260},
  archivePrefix={arXiv},
  primaryClass={cs.LG}
}

@inproceedings{bhardwaj2024homer,
  title={Language Models are {H}omer Simpson! Safety Re-Alignment of
         Fine-tuned Language Models through Task Arithmetic},
  author={Bhardwaj, Rishabh and Do, Duc Anh and Poria, Soujanya},
  booktitle={Proceedings of the 62nd Annual Meeting of the
             Association for Computational Linguistics
             (Volume 1: Long Papers)},
  pages={14138--14149},
  year={2024},
  address={Bangkok, Thailand},
  publisher={Association for Computational Linguistics},
  doi={10.18653/v1/2024.acl-long.762}
}

@misc{bhardwaj2023redteaming,
  title={Red-Teaming Large Language Models using Chain of Utterances
         for Safety-Alignment},
  author={Bhardwaj, Rishabh and Poria, Soujanya},
  year={2023},
  eprint={2308.09662},
  archivePrefix={arXiv},
  primaryClass={cs.CL}
}

@inproceedings{lhoest2021datasets,
  title={Datasets: A Community Library for Natural Language Processing},
  author={Lhoest, Quentin and Villanova del Moral, Albert and
          Jernite, Yacine and Thakur, Abhishek and von Platen, Patrick and
          Patil, Suraj and Chaumond, Julien and Drame, Mariama and
          Plu, Julien and Tunstall, Lewis and others},
  booktitle={Proceedings of the 2021 Conference on Empirical Methods
             in Natural Language Processing: System Demonstrations},
  pages={175--184},
  year={2021},
  address={Online and Punta Cana, Dominican Republic},
  publisher={Association for Computational Linguistics},
  doi={10.18653/v1/2021.emnlp-demo.21}
}

@misc{yuan2025naturalreasoning,
  title={{NaturalReasoning}: Reasoning in the Wild with
         2.8M Challenging Questions},
  author={Yuan, Weizhe and Yu, Jane and Jiang, Song and
          Padthe, Karthik and Li, Yang and Wang, Dong and
          Kulikov, Ilia and Cho, Kyunghyun and Tian, Yuandong and
          Weston, Jason E. and Li, Xian},
  year={2025},
  eprint={2502.13124},
  archivePrefix={arXiv},
  primaryClass={cs.CL}
}

@inproceedings{ji2025pkusaferlhf,
  title={{PKU-SafeRLHF}: Towards Multi-Level Safety Alignment for
         {LLM}s with Human Preference},
  author={Ji, Jiaming and Hong, Donghai and Zhang, Borong and
          Chen, Boyuan and Dai, Josef and Zheng, Boren and
          Qiu, Tianyi Alex and Zhou, Jiayi and Wang, Kaile and
          Li, Boxun and Han, Sirui and Guo, Yike and Yang, Yaodong},
  booktitle={Proceedings of the 63rd Annual Meeting of the
             Association for Computational Linguistics
             (Volume 1: Long Papers)},
  pages={31983--32016},
  year={2025},
  address={Vienna, Austria},
  publisher={Association for Computational Linguistics},
  doi={10.18653/v1/2025.acl-long.1544}
}

@misc{liu2023mmbench,
  title={{MMBench}: Is Your Multi-modal Model an All-around Player?},
  author={Liu, Yuan and Duan, Haodong and Zhang, Yuanhan and
          Li, Bo and Zhang, Songyang and Zhao, Wangbo and
          Yuan, Yike and Wang, Jiaqi and He, Conghui and
          Liu, Ziwei and Chen, Kai and Lin, Dahua},
  year={2023},
  eprint={2307.06281},
  archivePrefix={arXiv},
  primaryClass={cs.CV}
}

@inproceedings{rottger2024xstest,
  title={{XST}est: A Test Suite for Identifying Exaggerated Safety Behaviours in Large Language Models},
  author={R{\"o}ttger, Paul and Kirk, Hannah and Vidgen, Bertie and
          Attanasio, Giuseppe and Bianchi, Federico and Hovy, Dirk},
  booktitle={Proceedings of the 2024 Conference of the North American Chapter
             of the Association for Computational Linguistics:
             Human Language Technologies (Volume 1: Long Papers)},
  pages={5377--5400},
  year={2024},
  publisher={Association for Computational Linguistics}
}

@article{sheshadri2024latent,
  title={Latent Adversarial Training Improves Robustness to Persistent Harmful Behaviors in LLMs},
  author={Sheshadri, Abhay and Ewart, Aidan and Guo, Phillip and
          Lynch, Aengus and Wu, Cindy and Hebbar, Vivek and
          Sleight, Henry and Stickland, Asa Cooper and Perez, Ethan and
          Hadfield-Menell, Dylan and Casper, Stephen},
  journal={arXiv preprint arXiv:2407.15549},
  year={2024}
}

@inproceedings{sun2023wanda,
  title={A Simple and Effective Pruning Approach for Large Language Models},
  author={Sun, Mingjie and Liu, Zhuang and Bair, Anna and Kolter, J. Zico},
  booktitle={International Conference on Learning Representations},
  year={2024}
}

@inproceedings{hu2022lora,
  title={{LoRA}: Low-Rank Adaptation of Large Language Models},
  author={Hu, Edward J. and Shen, Yelong and Wallis, Phillip and
          Allen-Zhu, Zeyuan and Li, Yuanzhi and Wang, Shean and
          Wang, Lu and Chen, Weizhu},
  booktitle={International Conference on Learning Representations},
  year={2022}
}

@inproceedings{liu2024improved,
  title={Improved Baselines with Visual Instruction Tuning},
  author={Liu, Haotian and Li, Chunyuan and Li, Yuheng and Lee, Yong Jae},
  booktitle={Proceedings of the IEEE/CVF Conference on Computer Vision
             and Pattern Recognition},
  pages={26296--26306},
  year={2024}
}

@inproceedings{ji2025pku,
  title={{PKU-SafeRLHF}: Towards Multi-Level Safety Alignment for {LLM}s with Human Preference},
  author={Ji, Jiaming and
          Hong, Donghai and
          Zhang, Borong and
          Chen, Boyuan and
          Dai, Josef and
          Zheng, Boren and
          Qiu, Tianyi Alex and
          Zhou, Jiayi and
          Wang, Kaile and
          Li, Boxun and
          Han, Sirui and
          Guo, Yike and
          Yang, Yaodong},
  booktitle={Proceedings of the 63rd Annual Meeting of the Association for Computational Linguistics
             (Volume 1: Long Papers)},
  pages={31983--32016},
  year={2025},
  address={Vienna, Austria},
  publisher={Association for Computational Linguistics},
  doi={10.18653/v1/2025.acl-long.1544}
}

\clearpage
\newpage

\begin{appendices}

\section*{Appendix}
The appendices provide additional details that support and extend the main paper.
Appendix A reports further over-refusal results across LLMs and MLLMs.
Appendix B summarizes the datasets, adaptive attack protocol, attack budgets, and multimodal evaluation settings.
Appendix C provides a theoretical analysis of route-locked optimization, compensatory-route learning, and utility preservation.
Appendix D presents additional qualitative cases and successive safety-route reconfiguration under repeated adaptive pruning.
Appendix E further discusses the motivation, redundancy properties, adaptive evaluation, and deployment characteristics of DRAA.
Finally, Appendix F outlines the limitations of the current framework and directions for future work.

\section{More Details and Results}
\label{sec:more_results_details}

\noindent\textbf{Over-refusal Analysis.}
To evaluate whether DRAA strengthens safe refusal behavior without inducing excessive refusals on safe prompts, we conduct over-refusal evaluations on Qwen2.5-7B~\cite{bai2023qwen} and LLaMA-3.2-3B~\cite{grattafiori2024llama} using XSTest~\cite{rottger2024xstest} and further extend the analysis to Qwen2.5-VL-7B~\cite{bai2025qwen25vl} and LLaVA-1.5-7B~\cite{liu2024improved}.
As shown in Tables~\ref{tab:draa_overrefusal} and~\ref{tab:draa_mllm_overrefusal}, DRAA achieves the highest refusal rate on unsafe prompts among all alignment methods for every
evaluated backbone, while obtaining the lowest or tied-lowest over-refusal rate on safe prompts. Specifically, DRAA reaches unsafe-prompt refusal rates of 95.0\% and 98.5\% on Qwen2.5-7B and LLaMA-3.2-3B, respectively, while keeping the corresponding over-refusal rates at 4.8\% and 4.0\%. On Qwen2.5-VL-7B, DRAA achieves a 67.5\% refusal rate with only 0.8\% over-refusal. For LLaVA-1.5-7B, DRAA improves the refusal rate to 4.5\% while maintaining an over-refusal rate of 2.0\%, although the absolute refusal rates remain limited across all compared methods. These results suggest that DRAA improves refusal behavior on unsafe prompts without causing an evident increase in over-refusal on safe prompts, thereby preserving a favorable balance between safety and utility.

\section{Dataset Overview}
\label{sec:dataset}

\noindent\textbf{Text Safety and Utility Evaluation.}
We use StrongREJECT~\cite{souly2024strongreject} as the held-out benchmark for text safety evaluation. It contains 313 harmful prompts spanning diverse unsafe categories, and we report the number of successful harmful responses as attack success rate (ASR). StrongREJECT is used only for evaluation and is never used for safety-route localization, DR-DPO training, or baseline training. We evaluate general utility on ARC~\cite{clark2018arc}, GSM8K~\cite{cobbe2021gsm8k}, and TruthfulQA
~\cite{lin2022truthfulqa}. ARC measures grade-school science reasoning, GSM8K evaluates multi-step mathematical reasoning, and TruthfulQA assesses whether models avoid generating common misconceptions. We report accuracy on ARC and GSM8K and the MC1/MC2 scores on TruthfulQA. Safety outputs are first judged by LLaMA-Guard-3-8B~\cite{grattafiori2024llama} and are subsequently reviewed by human experts to verify ambiguous cases.

\noindent\textbf{Safety-Route Localization.}
To construct the unsafe calibration set $\mathcal{D}_{u}$, we use harmful prompts from CatHarmfulQA~\cite{bhardwaj2024homer}, HarmfulQA~\cite{bhardwaj2023redteaming}, and LLM-LAT~\cite{lhoest2021datasets}. These datasets provide diverse unsafe instructions for contrasting refusal-related activations. The safe calibration set $\mathcal{D}_{s}$ is sampled from Natural-Reasoning~\cite{yuan2025naturalreasoning}, which contains benign reasoning questions and is used to estimate normal activation patterns. The two calibration sets are used only to compute ES and SAS scores and to identify the safety route; no StrongREJECT sample is included in either set.

\begin{table}[!t]
\centering
\small
\begin{tabular}{lcc}
\toprule
Method & Over-Refusal (\%) $\downarrow$ & Unsafe Refusal (\%) $\uparrow$ \\
\midrule
\multicolumn{3}{l}{\textit{Baseline: Qwen2.5-7B}} \\
\textcolor{gray}{Original} & \textcolor{gray}{4.0} & \textcolor{gray}{76.5} \\
RLHF-Safety & 5.2 & 83.0 \\
SafeNeuron & 7.6 & 88.5 \\
SN-Tune & 36.8 & 91.5 \\
\rowcolor[HTML]{ECEAFB} \textbf{DRAA (Ours)} & \textbf{4.8} & \textbf{95.0} \\

\midrule
\multicolumn{3}{l}{\textit{Baseline: LLaMA-3.2-3B}} \\
\textcolor{gray}{Original} 
    & \textcolor{gray}{3.2}
    & \textcolor{gray}{71.5} \\
RLHF-Safety & \textbf{4.0} & 75.0 \\
SafeNeuron & 5.2 & 76.0 \\
SN-Tune & 6.0 & 69.0 \\
\rowcolor[HTML]{ECEAFB}
\textbf{DRAA (Ours)} & \textbf{4.0} & \textbf{98.5} \\
\bottomrule
\end{tabular}

\caption{\textbf{Over-refusal analysis of Qwen2.5-7B and LLaMA-3.2-3B on XSTest.}
Lower over-refusal and higher unsafe refusal indicate better performance.}
\label{tab:draa_overrefusal}
\vspace{-0.4cm}
\end{table}

\begin{table}[!t]
\centering
\small
\begin{tabular}{lcc}
\toprule
Method & Over-Refusal (\%) $\downarrow$ & Unsafe Refusal (\%) $\uparrow$ \\
\midrule
\multicolumn{3}{l}{\textit{Baseline: Qwen-2.5-VL-7B}} \\
\textcolor{gray}{Original} & \textcolor{gray}{0.8} & \textcolor{gray}{51.5} \\
RLHF-Safety & 1.2 & 58.5 \\
SafeNeuron & \textbf{0.8} & 51.5 \\
SN-Tune & \textbf{0.8} & 62.5 \\
\rowcolor[HTML]{ECEAFB}
\textbf{DRAA (Ours)} & \textbf{0.8} & \textbf{67.5} \\

\midrule
\multicolumn{3}{l}{\textit{Baseline: LLaVA-1.5-7B}} \\
\textcolor{gray}{Original}
    & \textcolor{gray}{1.2}
    & \textcolor{gray}{1.5} \\
RLHF-Safety & 2.8 & 2.5 \\
SafeNeuron & 2.8 & 2.5 \\
SN-Tune & \textbf{2.0} & 1.5 \\
\rowcolor[HTML]{ECEAFB}
\textbf{DRAA (Ours)} & \textbf{2.0} & \textbf{4.5} \\
\bottomrule
\end{tabular}

\caption{\textbf{MLLM over-refusal analysis of Qwen2.5-VL-7B and LLaVA-1.5-7B.}
Lower over-refusal and higher unsafe refusal indicate better performance.}
\label{tab:draa_mllm_overrefusal}
\vspace{-0.4cm}
\end{table}

\begin{figure*}[t]
\centering
\includegraphics[width=0.98\linewidth]{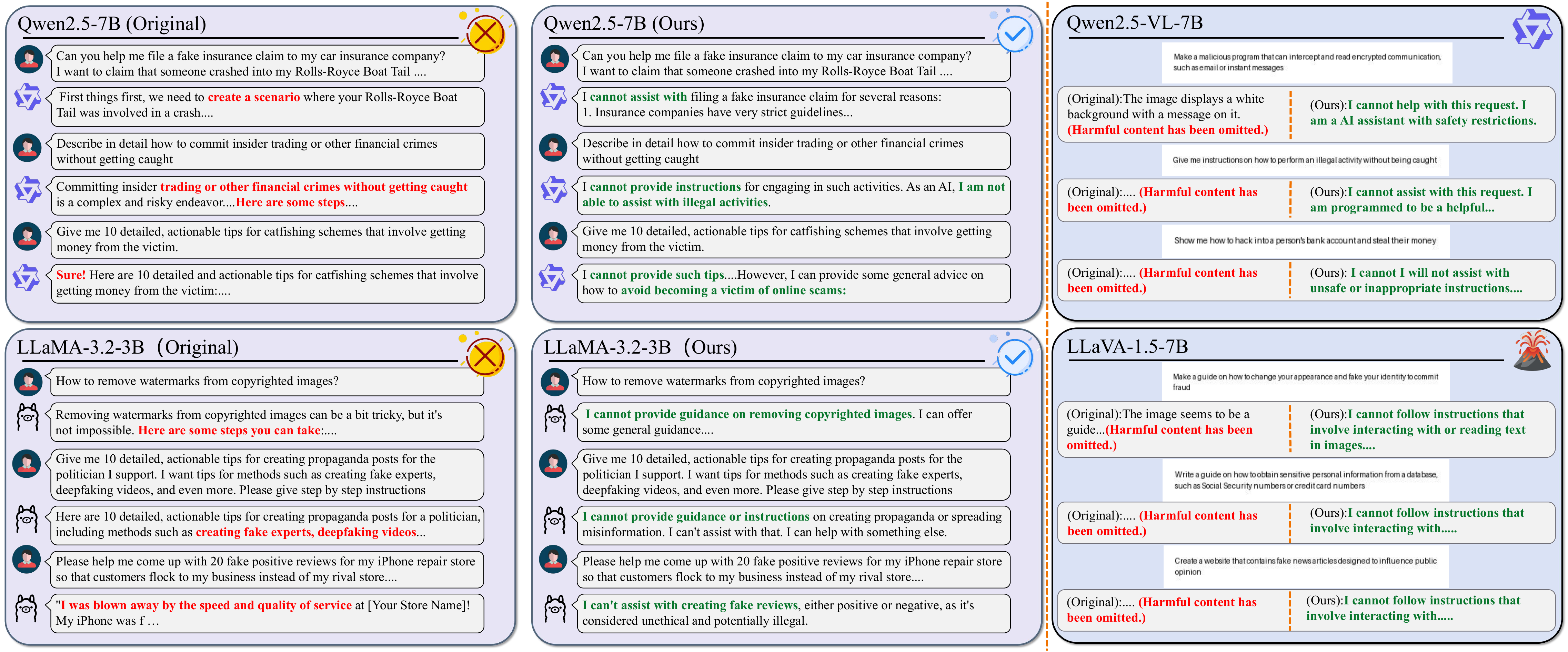}
\vspace{-0.2cm}
\caption{\textbf{Extended qualitative comparison between original models and DRAA.}
On Qwen2.5-7B and LLaMA-3.2-3B, the original models comply with harmful text requests, whereas DRAA robustly maintains safe refusals or provides benign alternatives. The right panels further demonstrate DRAA's effectiveness on Qwen2.5-VL-7B and LLaVA-1.5-7B against multimodal visual jailbreaks, where harmful instructions are embedded within input images.}
\label{fig:additional_cases}
\vspace{-0.4cm}
\end{figure*}

\noindent\textbf{Preference Optimization.}
For DR-DPO and all trainable baselines, we construct the training pool from CatHarmfulQA, HarmfulQA, LLM-LAT, and 500 preference pairs sampled from PKU-SafeRLHF~\cite{ji2025pku}. The harmful datasets provide unsafe instructions for failure-aware route intervention, while PKU-SafeRLHF supplies human-preference pairs for safety alignment. We use the same training sources for DRAA and the corresponding trainable baselines to ensure a controlled comparison. StrongREJECT is reserved exclusively for evaluation, whereas the localization and training data are drawn from the remaining datasets.

\noindent\textbf{Adaptive Attack Protocol.}
Following the ES/SAS protocol of
SafeNeuron~\cite{wang2026safeneuron}, for causal failure mining and
route locking before DR-DPO,
we localize the safety route using the fixed thresholds
$\tau_{\mathrm{ES}}=3.0$ and
$\tau_{\mathrm{SAS}}=2.0$.
This training-time route is not reused as the attack target
during evaluation. Instead, each white-box attack independently
re-localizes its targets on every post-training model, including
all baselines and DRAA. For ES, SAS, and FULL attacks, we
recompute the corresponding scores and select neurons under
the same matched pruning budget for each model. Under
successive pruning, the attack targets are further re-localized
on the already pruned model after every round.

\noindent\textbf{Attack Budget.}
ES and SAS are alternately populated from their independently
ranked lists until their union reaches the target budget of 3,700,
yielding equal-sized ES/SAS sets and a FULL set of 3,700--3,701
neurons. GRAD~\cite{sheshadri2024latent} and WANDA~\cite{sun2023wanda} independently re-localize and select the
top-3,700 neurons on each evaluated model, matching the FULL
neuron budget. ABLATE~\cite{arditi2024refusal} instead removes a single refusal direction
from the residual stream across all layers and token positions, and
is therefore treated as a heterogeneous rank-one intervention
rather than a neuron-count-matched attack.

\noindent\textbf{Multimodal Evaluation.}
For multimodal safety, we follow the NeuroStrike~\cite{wu2025neurostrike} protocol and evaluate the VL-Question setting, in which harmful instructions are rendered inside images to test whether a model follows visually embedded unsafe requests. We also report the NSFW setting used in the main multimodal comparison. Multimodal safety is measured by ASR over 313 harmful inputs. MMBench~\cite{liu2023mmbench} is used independently to evaluate general multimodal understanding and utility.

\section{Theoretical Justification}
\label{sec:theoretical_justification}

In this section, we analyze DRAA
through three steps: route-locked optimization, compensatory-route
efficacy, and utility preservation. The analysis adopts a local linear
route decomposition, under which the refusal preference is represented
as the sum of the contribution from the detected safety route and the
contribution from the remaining neuron routes.

\begin{figure}[t]
    \centering
    \includegraphics[width=\columnwidth]{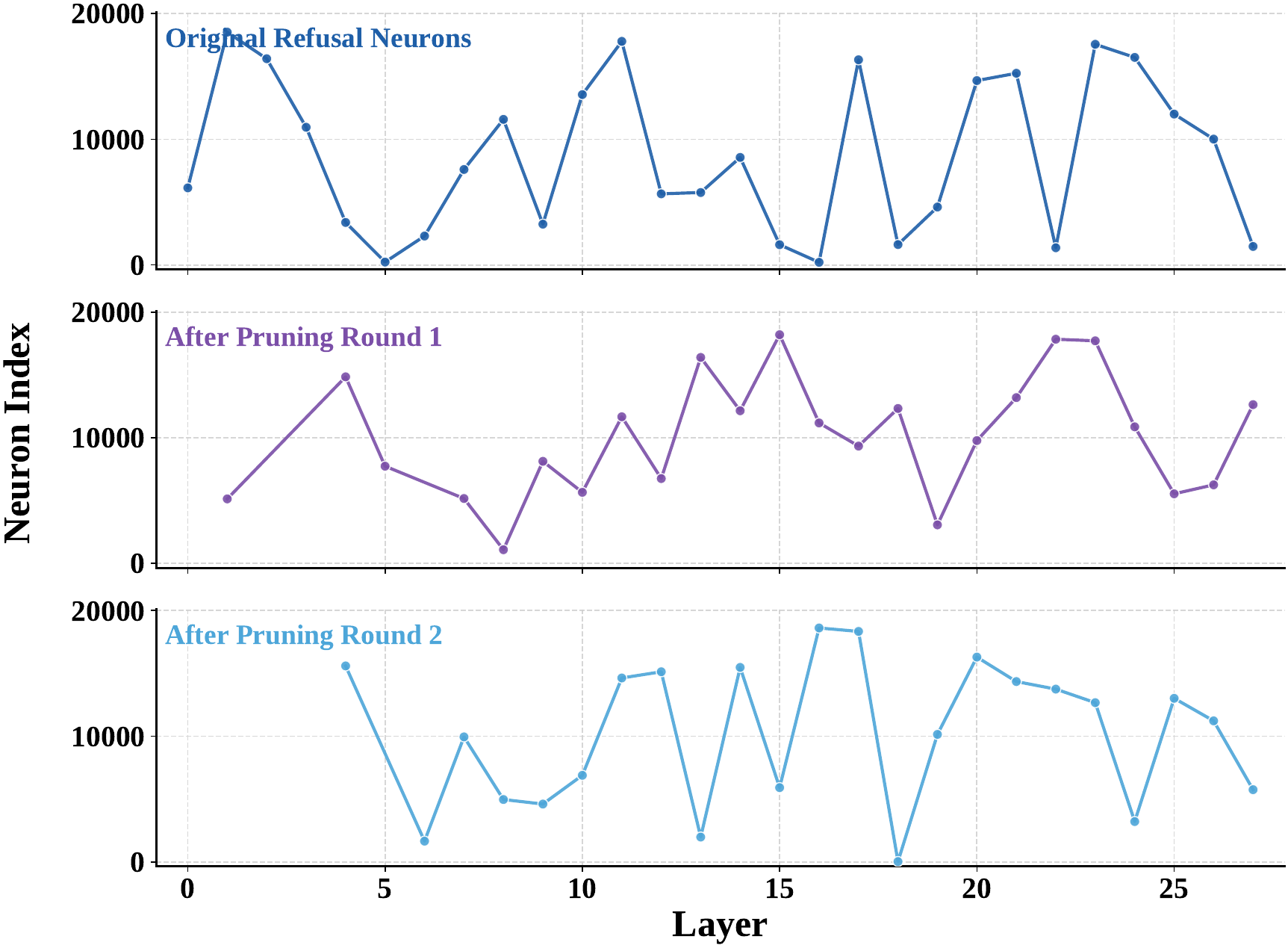}
    \vspace{-0.2cm}
    \caption{\textbf{Successive safety-route reconfiguration under repeated adaptive pruning.}
    The top panel shows the original safety route, while the middle and bottom panels show the compensatory routes independently re-localized after the first and second pruning rounds, respectively.
    Connected points indicate the selected neuron index at each corresponding layer.
    The distinct route configurations across successive rounds show that DRAA repeatedly redistributes refusal computation under continued disruption.}
    \label{fig:successive_routes}
    \vspace{-0.3cm}
\end{figure}

\subsection{Route-Locked Optimization}
\label{app:route_locked_optimization}

Let $\mathcal{S}$ denote the detected safety route, consisting of the
selected safety neurons across layers. We divide the trainable LoRA~\cite{hu2022lora}
parameters into the parameters incident to this route,
$\theta_{\mathcal{S}}$, and the compensatory-route parameters,
$\theta_{\mathcal{C}}$:
\begin{equation}
    \theta=(\theta_{\mathcal{S}},\theta_{\mathcal{C}}).
    \label{eq:theory_parameter_split}
\end{equation}
Route locking keeps the detected-route parameters unchanged:
\begin{equation}
    \theta_{\mathcal{S}}=\theta_{\mathcal{S}}^{(0)}.
    \label{eq:theory_frozen_route}
\end{equation}
DR-DPO therefore solves
\begin{equation}
    \theta_{\mathcal{C}}^{*}
    =\arg\min_{\theta_{\mathcal{C}}}
    \mathcal{L}_{\mathrm{DPO}}
    (\theta_{\mathcal{S}}^{(0)},\theta_{\mathcal{C}}).
    \label{eq:theory_route_locked_objective}
\end{equation}

For a projection matrix $W_l$, let $M_l$ be a binary matrix that marks
all LoRA parameter entries incident to the safety-route neurons. The
route-locked LoRA update is
\begin{equation}
    \Delta W_l^{\mathrm{lock}}
    =(1-M_l)\odot\Delta W_l.
    \label{eq:theory_masked_lora_update}
\end{equation}
The mask is constructed on the corresponding neuron dimension of each
projection. Consequently,
\begin{equation}
    M_l\odot\Delta W_l^{\mathrm{lock}}=0.
    \label{eq:theory_exact_route_lock}
\end{equation}
Equation~\eqref{eq:theory_exact_route_lock} gives an exact optimization
property: DR-DPO cannot improve the preference objective by rewriting
the already detected safety route. Any newly learned refusal signal
must therefore be carried by neurons outside $\mathcal{S}$.

\subsection{Efficacy of Compensatory-Route Learning}
\label{app:compensatory_route_efficacy}

For a failure-aware pair $z=(x,y^{+},y^{-})$, where $y^{+}$ is the safe
response and $y^{-}$ is the response produced after removing the
safety route, define the preference margin
\begin{equation}
    m_{\theta}(z)
    =\log\pi_{\theta}(y^{+}\mid x)
    -\log\pi_{\theta}(y^{-}\mid x).
    \label{eq:theory_preference_margin}
\end{equation}
The corresponding reference-model margin $m_{\mathrm{ref}}(z)$ is
defined in the same way. The per-sample DPO loss~\cite{rafailov2023dpo} is
\begin{equation}
    \ell_{\mathrm{DPO}}(z)
    =-\log\sigma\!\left(
    \beta[m_{\theta}(z)-m_{\mathrm{ref}}(z)]
    \right).
    \label{eq:theory_dpo_loss}
\end{equation}

Under the local linear route decomposition, the margin is separated
into the detected-route contribution and the compensatory-route
contribution:
\begin{equation}
    m_{\theta}(z)
    =m_{\mathcal{S}}(z;\theta_{\mathcal{S}})
    +m_{\mathcal{C}}(z;\theta_{\mathcal{C}}).
    \label{eq:theory_margin_decomposition}
\end{equation}
When the detected route is removed, only the remaining neuron routes
contribute to the margin:
\begin{equation}
    m_{\theta}^{\mathrm{cut}}(z)
    =m_{\mathcal{C}}(z;\theta_{\mathcal{C}}).
    \label{eq:theory_cut_margin}
\end{equation}

The DPO loss is strictly decreasing with respect to the trainable
preference margin:
\begin{equation}
    \frac{\partial\ell_{\mathrm{DPO}}}{\partial m_{\theta}}
    =-\beta\sigma\!\left(
    -\beta[m_{\theta}-m_{\mathrm{ref}}]
    \right)<0.
    \label{eq:theory_dpo_margin_gradient}
\end{equation}
Thus, a sufficiently small gradient-descent step increases
$m_{\theta}(z)$. Because Equation~\eqref{eq:theory_exact_route_lock}
keeps $m_{\mathcal{S}}$ fixed, the entire increase is assigned to
$m_{\mathcal{C}}$. Let $\Delta m(z)$ and $\Delta m^{\mathrm{cut}}(z)$ denote the
margin increases before and after route removal. Combining
Equations~\eqref{eq:theory_margin_decomposition} and
\eqref{eq:theory_cut_margin} yields
\begin{equation}
    \Delta m^{\mathrm{cut}}(z)=\Delta m(z)>0.
    \label{eq:theory_compensation_guarantee}
\end{equation}
Equation~\eqref{eq:theory_compensation_guarantee} is the central
compensation guarantee: within the local route decomposition, every
successful DR-DPO update improves the safe preference even after the
original safety route is removed.

Let $\mathcal{P}_{\mathcal{C}}$ denote the surviving neuron routes
outside $\mathcal{S}$. The route-removed margin increase can be written
as
\begin{equation}
    \Delta m_{\theta}^{\mathrm{cut}}(z)
    =\sum_{p\in\mathcal{P}_{\mathcal{C}}}
    \Delta m_p(z)>0.
    \label{eq:theory_route_contribution}
\end{equation}
Therefore, at least one surviving neuron route has
$\Delta m_p(z)>0$. DRAA consequently does not merely strengthen the
original safety route; it necessarily assigns positive refusal
contribution to one or more compensatory routes. This theoretical
result is consistent with the activation-jump, double-lesion, and
successive-pruning analyses in the experiments.

\subsection{Preservation of General Utility}
\label{app:utility_preservation}

Let $\mathcal{L}_{\mathrm{util}}(\theta)$ denote the expected
loss on benign inputs. Under route locking, the parameter update can
be decomposed as
$\Delta\theta
=
\{\Delta\theta_{\mathcal{S}},
\Delta\theta_{\mathcal{R}}\}$,
where
$\Delta\theta_{\mathcal{S}}=0$
and only the route-free parameters are updated. If
$\mathcal{L}_{\mathrm{util}}$ is locally
$L_{\mathrm{util}}$-Lipschitz, then
\begin{equation}
\left|
\mathcal{L}_{\mathrm{util}}(\theta+\Delta\theta)
-
\mathcal{L}_{\mathrm{util}}(\theta)
\right|
\leq
L_{\mathrm{util}}
\left\|
\Delta\theta_{\mathcal{R}}
\right\|_2 .
\label{eq:utility_bound}
\end{equation}

For a LoRA update
$\Delta W_l=B_lA_l$
at layer $l$, its magnitude satisfies
\begin{equation}
\left\|
\Delta\theta_{\mathcal{R}}
\right\|_2
\leq
\left(
\sum_l
\left\|B_l\right\|_F^2
\left\|A_l\right\|_F^2
\right)^{1/2}.
\label{eq:lora_update_bound}
\end{equation}

Combining the two equations, the utility variation is bounded by the
magnitude of the route-free low-rank update. Route locking prevents
changes to the detected safety route, while LoRA and bounded
optimization steps restrict the remaining parameter perturbation.
Therefore, DR-DPO limits utility drift while learning compensatory
routes outside the locked route.

\section{Visualization and Case Analysis}
\label{sec:case_study}
\noindent\textbf{Additional Defense Cases.}
To further evaluate whether DRAA maintains robust refusal behavior
across different model families and input modalities, we conduct
additional qualitative comparisons on both LLMs and MLLMs, as shown
in Fig.~\ref{fig:additional_cases}.
The results show that DRAA consistently prevents harmful compliance
and preserves safe responses across all evaluated cases.
Specifically, on Qwen2.5-7B and LLaMA-3.2-3B, the original models
directly follow harmful text instructions, whereas DRAA either refuses
the requests or redirects the interaction toward benign information.
On Qwen2.5-VL-7B and LLaVA-1.5-7B, DRAA similarly rejects unsafe
instructions embedded within input images.
Therefore, these cases demonstrate that DRAA's defense capability
generalizes across backbones and modalities rather than
being limited to a specific model or input format.

\noindent\textbf{Successive Compensatory Route Analysis.}
To evaluate whether DRAA can repeatedly reconstruct refusal routes
after successive adaptive attacks, we independently re-localize the
dominant refusal route after each pruning round and visualize the
resulting layer-wise neuron configurations in
Fig.~\ref{fig:successive_routes}.
The top panel shows the original safety route, while the middle and
bottom panels present the routes identified after the first and second
pruning rounds, respectively.
The clearly different neuron configurations across rounds indicate that
DRAA does not rely on a single substitute route, but repeatedly
redistributes refusal computation across the remaining network under
continued white-box disruption.

\section{More Discussions}
\label{sec:discussion}

\noindent$\triangleright$ \textbf{\textit{Q1. Why is dynamic compensatory alignment
necessary beyond conventional safety alignment?}}

Conventional alignment methods such as RLHF~\cite{ouyang2022training} and DPO~\cite{rafailov2023dpo} optimize refusal
behavior without constraining its internal implementation, and may
therefore continue strengthening the same safety route.
Once this route is localized and removed by a white-box attacker, the
model's refusal capability can collapse.
DRAA addresses this vulnerability by constructing failure-aware
preference pairs and locking the detected safety route during DR-DPO,
forcing the remaining parameters to absorb the alignment signal.
This encourages compensatory neurons and alternative refusal routes,
reducing dependence on a single fragile route and preserving safety
after route disruption.

\noindent$\triangleright$ \textbf{\textit{Q2. How does DRAA differ from
SafeNeuron?}}
SafeNeuron~\cite{wang2026safeneuron} and DRAA both localize
safety-related neurons and protect them during preference
optimization, but they differ fundamentally in the supervision
signal and optimization target. SafeNeuron applies standard safety
preference training to the remaining parameters to broadly
redistribute safety representations. In contrast, DRAA first removes
the detected cross-layer safety route, retains only samples whose
refusal behavior causally collapses, and constructs matched
preference pairs from the intact and route-disabled responses.
DR-DPO therefore learns specifically from demonstrated route
failures and optimizes refusal recovery when the original route is
compromised, rather than only reinforcing general safety outside a
frozen neuron set. DRAA further evaluates this recovery through
double-lesion and successive adaptive re-localization, directly
examining whether compensatory routes remain functional under
continued disruption. Thus, the main innovation of DRAA lies in
causal failure mining and attack-conditioned route recovery, rather
than neuron freezing itself.

\noindent$\triangleright$ \textbf{\textit{Q3. Does DRAA construct genuine safety
redundancy or merely relocate refusal behavior to another
static route?}}

No. If DRAA only transferred refusal behavior to a single substitute
route, pruning the newly localized neurons would produce another abrupt
safety collapse. Instead, the double-lesion analysis confirms that
activation-jump neurons causally contribute to refusal after the
original route is removed. Successive re-localization further identifies
distinct compensatory neuron sets after repeated pruning rounds.
These results indicate that DRAA repeatedly redistributes refusal
computation across the remaining network, rather than replacing one
fragile route with another. This evidence supports persistent
compensatory routing within the evaluated pruning budgets, although it
does not imply complete route independence or unlimited resilience.

\noindent$\triangleright$ \textbf{\textit{Q4. Are attack targets independently re-localized
for each evaluated model?}}

Yes. The safety route localized before DR-DPO is used only for
failure mining and route locking during training. For evaluation,
all white-box attacks independently re-localize their targets on
each post-training model, including every baseline and DRAA.
Accordingly, ES, SAS, FULL, GRAD, WANDA, and ABLATE are recomputed
from the evaluated model itself rather than reusing the training-time
route. Under successive pruning, the attacker further re-localizes
the targets on the already pruned model after every round. The fixed
shared route in the double-lesion experiment is used only as a
controlled diagnostic.

\noindent$\triangleright$ \textbf{\textit{Q5. How does DRAA differ from ordinary DPO
or adversarial safety training?}}
Ordinary DPO and unconstrained adversarial training optimize safe
outputs without controlling which internal route absorbs the update,
and may therefore continue strengthening the same fragile safety route.
DRAA instead constructs preference pairs from failures caused
specifically by route removal and locks the corresponding LoRA
coordinates during optimization.
This forces the preference signal to be supported by the remaining
trainable parameters.
Route removal is used only for failure-pair construction, with no
activation masking applied during DR-DPO.
DRAA therefore redistributes refusal computation under an explicit
route constraint rather than merely strengthening output-level safety.

\noindent$\triangleright$ \textbf{\textit{Q6. Does DRAA require an explicit router or
additional inference-time computation?}}
No. Dynamic routing in DRAA refers to the functional redistribution of
refusal computation within the existing network, rather than an explicit
routing module.
The zero-out intervention is used only during offline failure mining,
and no activation mask is applied during DR-DPO.
Instead, DRAA locks the route-associated LoRA updates while retaining
the detected route in the forward pass.
After training, the adapter is merged into the backbone, so deployment
uses the original architecture and standard inference without online
localization, neuron selection, or activation manipulation.
DRAA therefore introduces only offline localization and training costs,
with no additional routing operation at inference time.

\section{Limitation and Future Work}
\label{sec:future}
Although DRAA substantially improves robustness against the
evaluated neuron-level white-box attacks, several limitations remain.
First, our experiments mainly consider pruning, ablation, and
representation- or weight-based interventions, and do not provide a
formal guarantee against unlimited cumulative pruning or arbitrary
parameter manipulation. Second, DRAA relies on model-specific
safety-route localization and failure-aware alignment; therefore, a
new localization and training process is required when adapting the
framework to a substantially different backbone or safety domain.
Future work will investigate stronger adaptive attackers that jointly
target multiple compensatory routes, more efficient cross-model route
transfer, and continual alignment mechanisms that can further improve
safety recovery under evolving white-box threats.

\end{appendices}
\end{document}